\documentclass[12pt]{article} 
\usepackage{epsfig,amsmath,amssymb,graphics,color,calc} 
\usepackage{cite}

\usepackage{amsmath} 
\usepackage{amsthm}
\usepackage{amssymb}
\usepackage{amsopn}
\newcommand{\sezione}[2]{ 
\refstepcounter{section}\label{#2} 
\setcounter{equation}{0} 
\setcounter{subsection}{0} 
\addcontentsline{toc}{section} 
      {\normalsize\textbf{\thesection.\ #1}} 
\bigskip\bigskip\noindent 
\normalsize\textbf{\thesection.\ #1}\nopagebreak\smallskip\nopagebreak} 
\def\thesection{{\normalsize\arabic{section}}} 
\newcommand{\subsec}[2]{ 
\refstepcounter{subsection}\label{#2} 
\addcontentsline{toc}{subsection} 
      {\normalsize\normalfont\textit{\thesubsection.\ #1}} 
\medskip\medskip\noindent 
\normalsize\normalfont
\textit{\thesubsection. \ #1}\nopagebreak\smallskip\nopagebreak} 
\def\thesubsection{{\normalsize
{\arabic{section}.\arabic{subsection}}}} 
\newcounter{appendice}

\newtheorem{teo}{Theorem}[section]	\newtheorem{pro}[teo]{Proposition}
\newtheorem{defi}[teo]{Definition}	\newtheorem{lem}[teo]{Lemma}
\newtheorem{cor}[teo]{Corollary}	\newtheorem{rem}[teo]{Remark}
\newtheorem{con}[teo]{Condition}

\newcommand{\bteo}[1]{\begin{teo}\label{#1}}
\newcommand{\bpro}[1]{\begin{pro}\label{#1}}
\newcommand{\bdefi}[1]{\begin{defi}\label{#1}}
\newcommand{\blem}[1]{\begin{lem}\label{#1}}
\newcommand{\bcor}[1]{\begin{cor}\label{#1}}
\newcommand{\brem}[1]{\begin{rem}\label{#1}}
\newcommand{\bcon}[1]{\begin{con}\label{#1}}

\newcommand{\eteo}{\end{teo}}	\newcommand{\epro}{\end{pro}}
\newcommand{\edefi}{\end{defi}}	\newcommand{\elem}{\end{lem}}
\newcommand{\ecor}{\end{cor}}	\newcommand{\erem}{\end{rem}}
\newcommand{\econ}{\end{con}}

\newcommand{\Pro}{\noindent{\it Proof.\/}\ \ }
\newcommand{\Rem}{\noindent{\it Remark.\/}\ \ }
\renewcommand{\eqref}[1]{(\ref{#1})}

\newcommand{\be}[1]{\begin{equation}\label{#1}}
\newcommand{\bea}[1]{\begin{eqnarray}\label{#1}}
\newcommand{\besn}{\begin{equation*}}
\newcommand{\beasn}{\begin{eqnarray*}}
\newcommand{\nn}{\nonumber}
\renewcommand{\(}{\left(}		\renewcommand{\)}{\right)}
\renewcommand{\[}{\left[}		\renewcommand{\]}{\right]}
\renewcommand{\lg}{\left\{}		\newcommand{\rg}{\right\}}
\newcommand{\lmo}{\left|}		\newcommand{\rmo}{\right|}

\newcommand{\su}{\subset}	\newcommand{\ssu}{\subset\subset}
\newcommand{\sm}{\setminus}	\newcommand{\es}{\emptyset}

\newcommand{\dis}{\mathop{\rm d}\nolimits}
\newcommand{\supp}{\mathop{\rm supp}\nolimits}

	\newcommand{\id}{{1 \mskip -5mu {\rm I}}}
 
\newcommand{\noi}{\noindent}

\newcommand{\ul}[1]{\underline{#1}}

\renewcommand{\a}{\alpha}	\renewcommand{\b}{\beta}	
			
\newcommand{\e}{\varepsilon}	\newcommand{\f}{\varphi} 
\newcommand{\g}{\gamma}		
\renewcommand{\k}{\varkappa}

\newcommand{\s}{\sigma}		
\renewcommand{\t}{\tau}		
	
\newcommand{\z}{\zeta}
		\newcommand{\F}{\Phi}
\newcommand{\G}{\Gamma}		\renewcommand{\L}{\Lambda}
	\renewcommand{\O}{\Omega} 
		\newcommand{\Ps}{\Psi}

	\newcommand{\cF}{\mathcal F} 
\newcommand{\cG}{\mathcal G}	\newcommand{\cH}{\mathcal H} 
	 
	\newcommand{\cL}{\mathcal L} 
\newcommand{\cM}{\mathcal M}	\newcommand{\cN}{\mathcal N} 
	 
\newcommand{\cQ}{\mathcal Q}	\newcommand{\cR}{\mathcal R} 
\newcommand{\cS}{\mathcal S}	\newcommand{\cT}{\mathcal T} 
	 
	\newcommand{\cX}{\mathcal X} 
	 
	\newcommand{\bB}{\mathbb B} 
	 
\newcommand{\bE}{\mathbb E}	 
\newcommand{\bG}{\mathbb G}	 
	 
	\newcommand{\bL}{\mathbb L} 
	\newcommand{\bN}{\mathbb N} 
	 
	\newcommand{\bR}{\mathbb R} 
	\newcommand{\bT}{\mathbb T} 
	\newcommand{\bV}{\mathbb V} 
	 
	\newcommand{\bZ}{\mathbb Z} 
	\newcommand{\Z}{\mathbb Z}
\newcommand{\newatop}[2]{\genfrac{}{}{0pt}{}{#1}{#2}} 
 
\newcommand{\Es}{Y} 
\newcommand{\dEs}{y} 
\newcommand{\connj}{\stackrel{j}{\longleftrightarrow}}

\newcommand{\inc}{\textrm{inc}} 
 
\newcommand{\compa}{\textrm{comp}} 
 
\newcommand{\rest}{\!\restriction\!} 
\newcommand{\env}{\cQ} 
\newcommand{\clos}[1]{\overline{#1}} 
\newcommand{\proj}[1]{\widehat{#1}} 
\newcommand{\tree}{\bT} 
\renewcommand{\hat}{\widehat} 
\renewcommand{\tilde}{\widetilde} 
\newcommand{\disuno}{\mathrm d_1} 
\renewcommand{\dis}{\disuno} 
\newcommand{\disinfinito}{\mathrm d_\infty} 
\newcommand{\diamuno}{\mathrm{diam}_1} 
 
\newcommand{\diaminfinito}{\mathrm{diam}_\infty} 
\renewcommand{\complement}{\mathrm{c}} 
\newcommand{\incr}{2.2} 
\newcommand{\iincr}{3.2} 

\definecolor{light}{gray}{.9}

\begin{document} 
\begin{titlepage} 
%
%
%
\par\vskip 1cm\vskip 2em 
 
\begin{center} 
 
{\LARGE Graded cluster expansion for lattice systems} 
\par 
\vskip 2.5em \lineskip .5em 
{\large 
\begin{tabular}[t]{c} 
$\mbox{Lorenzo Bertini}^{1} \phantom{m} \mbox{Emilio N.M.\ Cirillo}^{2} 
\phantom{m} \mbox{Enzo Olivieri}^{3}$ 
\\ 
\end{tabular} 
\par 
} 
 
\medskip 
{\small 
\begin{tabular}[t]{ll} 
{\bf 1} & {\it 
Dipartimento di Matematica, Universit\`a di Roma La Sapienza}\\ 
&  Piazzale Aldo Moro 2, 00185 Roma, Italy\\ 
&  E--mail: {\tt bertini@mat.uniroma1.it}\\ 
\\ 
{\bf 2} & {\it 
Dipartimento Me.\ Mo.\ Mat., Universit\`a di Roma La Sapienza}\\ 
&  Via A.\ Scarpa 16, 00161 Roma, Italy\\ 
&  E--mail: {\tt cirillo@dmmm.uniroma1.it}\\ 
\\ 
{\bf 3} & {\it 
Dipartimento di Matematica, Universit\`a di Roma Tor Vergata}\\ 
& Via della Ricerca Scientifica, 00133 Roma, Italy\\ 
& E--mail: {\tt olivieri@mat.uniroma2.it}\\ 
\end{tabular} 
} 
\bigskip 
\end{center} 
\par\noindent 
{\bf Communicating author:} Enzo Olivieri\\ 
{\bf E\_mail:} \texttt{olivieri@mat.uniroma2.it}\\ 
{\bf Telephone number:} +39--06--72594686\\ 
{\bf Fax number:} +39--06--72594699 
\vskip 1 em 
 
\centerline{\bf Abstract} 
\smallskip 
In this paper we develop a general theory which provides  
a unified treatment of two apparently different problems.  
The weak Gibbs property of measures 
arising from the application of Renormalization Group maps  
and the mixing properties of disordered 
lattice systems in the Griffiths' phase.   
We suppose that the system satisfies a mixing condition 
in a subset of the lattice whose complement is sparse enough 
namely, large regions are widely separated.  
We then show how it is possible to construct a convergent multi-scale  
cluster expansion.

\vskip 0.8 em 
 
\vfill 
 
 
\bigskip\bigskip 
\footnoterule 
\vskip 1.0em 
{\small 
\noindent 
The authors acknowledge the support of  
Cofinanziamento MIUR. 
\vskip 1.0em 
\noindent 
} 
\end{titlepage} 
\vfill\eject 
 
\sezione{Introduction}{s:int} 
\par\noindent 
In this paper we develop a general theory which provides  
a unified treatment of two 
apparently different problems: (i) the weak Gibbs property of measures 
arising from the application of Renormalization Group (RG) maps to Gibbs 
states of lattice systems and (ii) the  mixing properties of disordered 
lattice systems in the so called Griffiths' phase.  Let us explain the 
main features of these issues. 
 
\subsec{Weak Gibbsianity of renormalized measures}{s:grm} 
\par\noindent 
Renormalization group is a fundamental method in modern 
theoretical physics. It has  been originally introduced to analyze 
scale invariant situations that are typical of statistical mechanical systems 
at their critical point. However it also exhibits its power for non--critical 
systems that deserve to be analyzed on appropriate scales, 
see \cite{[BK],[GaKK]}. 
The RG maps are defined as follows. 
 
Consider a $d$--dimensional 
lattice spin system (object system) whose state space is 
$\cX:=\bigotimes_{x\in\cL}\cX_x$, where $\cL:=\bZ^d$ and each  
$\cX_x$ is a copy of the same finite set $\cX_0$. 
We set 
$\cL^{(\ell)}:=\ell\bZ^d$, with $\ell\in\bN$, and  partition 
$\cL$ as the disjoint union of $\ell$--boxes  
$\cL=\bigcup_{i\in\cL^{(\ell)}}Q_\ell(i)$ where 
$Q_\ell(i)$ 
is the cube of side length $\ell$ with $i$ the site with smallest 
coordinates. 
Moreover, 
to each $i\in\cL^{(\ell)}$ we associate a 
{\it renormalized spin} $m_i$ taking value in a finite state space  
$\cM^{(\ell)}_i$, 
with each $\cM^{(\ell)}_i$ a copy of the same finite set $\cM^{(\ell)}_0$. 
We assign the normalized non--negative kernel 
$T_\ell(\s_{Q_\ell(i)},m_i)$ with 
$\s_{Q_\ell(i)}\in\bigotimes_{x\in Q_\ell(i)}\cX_x$ and $m_i\in\cM^{(\ell)}_i$. 
Given a Gibbs measure (w.r.t.\ an absolutely summable potential, for instance 
a finite range potential) 
$\mu$ on $\cX$, the renormalized measure $\nu^{(\ell)}$ on the renormalized  
space $\cM^{(\ell)}=\bigotimes_{i\in\cL^{(\ell)}}\cM^{(\ell)}_i$ is defined by 
its finite dimensional distributions  
$$ 
\nu^{(\ell)}(M^{(\ell)}:\,M^{(\ell)}_V=m_V) 
      =\sum_{\sigma\in\cX_{\Lambda}}\mu(\eta:\,\eta_\Lambda=\sigma) 
       \prod_{i\in V}T_\ell(\sigma_{Q_\ell(i)},m_i) 
$$ 
where $V$ is a finite subset of $\cL^{(\ell)}$, 
$\Lambda:=\bigcup_{i\in V} Q_\ell(i)$,  
$\cX_\Lambda:=\bigotimes_{x\in\Lambda}\cX_x$, and 
$m_V\in\cM^{(\ell)}_V:=\bigotimes_{i\in V}\cM^{(\ell)}_i$. 
We shall write $\nu^{(\ell)}=T_\ell\mu$. For the usual choices of the kernel, 
the semigroup property holds, namely 
$T_\ell T_{\ell'}=T_{\ell\ell'}$. 
 
An easy example is the {\it decimation} transformation where 
$\cM^{(\ell)}_i=\cX_i$, for all $i\in\cL^{(\ell)}$, 
and 
$T_\ell^{\mathrm{dec}}(\s_{Q_\ell(i)},m_i) 
 =\delta(\sigma_i-m_i)$; 
$m_i$, with $i\in\cL^{(\ell)}$, are the ``surviving spins." 
Another important 
example is the Block Averaging Transformation (BAT) that we discuss  
in the case $\cX_0:=\{-1,+1\}$:  
for each $i\in\cL^{(\ell)}$ the single renormalized spin 
configuration space is 
$\cM^{(\ell)}_i=\{-\ell^d,-\ell^d+2,\dots,+\ell^ d\}$ and 
$T_\ell^{\textrm{bat}}(\s_{Q_\ell(i)},m_i) 
 =\delta\Big(\sum_{x\in Q_\ell(i)}\s_x-m_i\Big)$. 
 
Theoretical reasons and many applications lead us to analyze the map on the 
potentials induced by the map $T_\ell$ that was defined on infinite volume 
Gibbs measures. 
A preliminary condition is that the renormalized measure is Gibbsian in the 
Dobrushin--Lanford--Ruelle sense i.e., its conditional probabilities have the 
Gibbs form with respect to an absolutely summable potential that we call 
renormalized potential \cite{[CasG],[GK]}. 
Another, hopefully equivalent, approach consists in defining at finite 
volume a 
map acting directly on the Hamiltonians in the following way. 
Given a box $V\subset\subset\cL^{(\ell)}$, let 
$\Lambda:=\bigcup_{i\in V}Q_\ell(i)$ be the 
corresponding box in $\cL$,   
$\cX_{\Lambda}:=\cX_0^\Lambda=\{-1,+1\}^\Lambda$,  
and $-H_\Lambda$ the energy of the object system, 
we write 
$$ 
e^{+H^{(\ell)}_V(m)}=\sum_{\s\in\cX_{\L}} 
                           e^{+H_\L(\s)}\prod_{i\in V} 
                           T_\ell(\s_{Q_\ell(i)},m_i) 
$$ 
where we have included in $H$ the inverse temperature. 
Now the problem is to extract the potential from the renormalized Hamiltonian 
$H^{(\ell)}_V$  
via a procedure still having a sense in the thermodynamic limit. 
For this purpose a crucial role is played by the so called {\em constrained 
systems} i.e., the object system conditioned on some fixed renormalized spin 
configuration. 
Given a renormalized spin configuration $m\in\cM^{(\ell)}_V$, 
the constrained measure in $\Lambda$ is defined by: 
$$ 
\mu_{m,\L}^{(\ell)}(\sigma)= 
\frac{{\displaystyle 
       e^{+H_\L(\s)} 
       \prod_{i\in V}T_\ell(\s_{Q_\ell(i)},m_i) 
      }} 
     {{\displaystyle 
       \sum_{\s\in\cX_{\L}}e^{+H_\L(\s)} 
       \prod_{i\in V}T_\ell(\s_{Q_\ell(i)},m_i) 
      }} 
$$ 
Note that the configuration $m$ plays the role of a parameter. 
In the case of the decimation, 
$\mu_{m,\L}^{(\ell)}$ is nothing but the original Gibbs 
measure in the volume obtained by removing from $\Lambda$ 
the set $V$ of the surviving sites, 
{\it conditioned} to the configuration $m$ on $V$. 
 
As it has been shown in many examples, see \cite{[EFS]}, it may happen that 
$\nu^{(\ell)}$ is 
not Gibbsian. Typically this pathology manifests itself as violation of {\it 
quasi--locality}, a necessary condition for Gibbsianity. 
More precisely, quasi--locality is a continuity property of the conditional 
probabilities consisting in a weak dependence 
on very far conditioning spins, see \cite{[EFS]} for more details. 
This violation of quasi--locality, in turn, is often a consequence of a first 
order phase transition of a constrained model corresponding to a  
particular renormalized configuration $m$. 
On the positive side, to avoid the pathology of non--Gibbsianity, we need 
absence of phase transitions, in 
a very strong sense, of the constrained model for {\em all} 
possible values of $m$. For instance 
when the object system is in the high temperature regime, 
the usual perturbative expansion for the 
constrained models is sufficient to compute the renormalized potentials, 
see \cite{[Cam],[GP],[I]}. However, in order to 
get close to the critical point, certainly we have to use other, more 
powerful, perturbative theories. 
 
\medskip 
We discuss, now, these different perturbative 
theories in the concrete case of systems above their critical 
temperature $T_c$. 
Usual high temperature expansions work only for temperatures $T$ sufficiently 
larger than $T_c$; they basically involve perturbations around a universal 
reference system consisting of independent spins: all interactions are 
expanded treating every lattice system in the same manner. 
In \cite{[O],[OP]} another perturbative expansion has been 
introduced, around a non trivial model--dependent reference system, that we 
call {\it scale--adapted expansion}. The small parameter is no more 
$\b=1/T$ but, rather, the ratio between the correlation length 
(at the given temperature $T>T_c$) and the length scale $L$ at which we 
analyze our system. The geometrical objects (polymers) involved in 
the scale--adapted expansion live on scale length $L$ whereas in the usual 
expansions they live on scale 1. 
Of course the smaller is $T-T_c$ the larger has to be taken the length $L$. 
 
A similar situation occurs for low temperature Ising ferromagnets at 
arbitrarily small  but non zero magnetic field $h$. Now we have 
another characteristic length, beyond the correlation length, the 
{\it critical length} which is of order $1/h$; it represents the minimal 
size of a droplet whose growth is energetically favorable and, at the same 
time, the minimal length required to {\it screen} the effect of a boundary 
condition opposite to the field. Thus in the part of the bulk far apart 
from the boundary more than the critical length, we see, 
uniformly in the boundary 
condition, the unique phase with magnetization parallel to the field. 
Also in this case of low temperature and not vanishing magnetic field we 
have to look at our system on a scale length $L$ sufficiently larger than 
the critical length (at low temperature and $h\neq0$ the correlation length is 
of order one). 
 
The scale--adapted expansions are based on a suitable {\it finite size 
condition} saying, roughly speaking, that if we look at the Gibbs 
measure in a box of sufficiently large side length $L$, then, uniformly in 
the boundary conditions, the correlations between 
observables localized at distance of order $L$ are smaller than 
$L^{-2(d-1)}$. It is 
proven in \cite{[O],[OP]}, that, 
for general short range lattice systems, assuming 
this finite size condition, it is possible to construct a convergent cluster expansion 
implying, in particular, exponential decay of correlations for any 
volume $\L$ (finite or infinite) given as disjoint union of $L$--boxes, with a 
decay rate independent of $\L$. 
We call this decay property {\it strong mixing}. Such a property 
implies uniqueness of the infinite volume Gibbs measure, 
we refer to \cite{[O],[OP],[MO2]} for more details. 
In some cases, like the two--dimensional standard Ising model,
strong mixing has been proven in the whole uniqueness region \cite{[MOS],[ScS]}. 
 
Starting from a strong mixing condition for the measure 
$\mu_{m}^{(\ell)}$, {\it uniform} in the renormalized configuration $m$, 
it is possible to prove, using the scale--adapted perturbative expansion, 
the Gibbsianity of $\nu^{(\ell)}$ by 
explicitly computing the renormalized potentials as convergent series, 
\cite{[HK],[BCO]}. In the particular case of BAT  
the condition on the constrained model 
can be deduced from a strong mixing property of the original model 
by using a strong form of the equivalence of ensembles \cite{[BCO],[CaM]}. 
 
\medskip 
The philosophy behind the use of scale--adapted expansions to study RG maps is 
that one first fixes the thermodynamic parameters and consequently chooses the 
renormalization scale $\ell$. 
It may happen that for a given scale $\ell$ and for particular 
values of the thermodynamic parameters 
a RG map is ill defined  but, keeping fixed the thermodynamic 
parameters, provided  one chooses a larger renormalization scale $\ell$, the 
pathology is removed. This is exactly the case when the decimation 
transformation is applied to a two--dimensional 
standard Ising model away from the coexistence line. 
In \cite{[EFS]} it is proven that for any decimation scale 
$\ell$ there exist values of $\b$ and $h$ such that the renormalized 
measure $\nu^{(\ell),\textrm{dec}}_{\b,h}$ 
is not Gibbsian. 
On the other hand, as shown in \cite{[MOdec]}, given $\beta$ and $h$ 
inducing the pathology for $\ell$, the renormalized measure 
$\nu^{(\bar\ell),\textrm{dec}}_{\beta,h}$ is 
Gibbsian provided the scale $\bar\ell$ is chosen sufficiently large.
It is also shown that the renormalized potential converges to zero as 
$\bar\ell$ goes to infinity. 
In \cite{[BCO]} this philosophy has been embraced to analyze  
the block averaging transformation on scale $\ell$.
In particular, for the two--dimensional 
standard Ising model at {\it any} $T>T_c$ and arbitrary $h$, the 
Gibbsianity of the renormalized measure 
$\nu^{(\ell),\textrm{bat}}_{\beta,h}$ is proven for $\ell$ 
{\it large enough}. Moreover the 
renormalized potential converges, in a suitable sense, 
to the expected trivial fixed point as $\ell$ goes to infinity. 
In order to perturbatively study convergence properties of  
the iterates of the renormalization group maps, even far from criticality, 
the use of scale--adapted expansions on increasing scales appears  
therefore very natural. 
 
We mention that there is a stronger notion of 
strong mixing, called {\it complete analyticity},
originally introduced for general short range lattice systems
by Dobrushin and Shlosman in \cite{[DS1],[DS2]} before \cite{[O],[OP]}.  
It consists of  
the exponential decay of correlations for {\it all} finite or 
infinite domains $\L$ (of arbitrary shape). 
Dobrushin and Shlosman also developed a finite size condition that involves
{\it all} the possible subsets of a given sufficiently large box and not
just the box itself like \cite{[MO2]}.
We emphasize that in their approach there is no minimal scale length.
On the other hand, the scale--adapted perturbative 
theory gives rise to a notion that has been called {\it restricted 
complete analyticity} or {\it complete analyticity for regular 
domains}; here ``regular" means ``multiple" of a sufficiently large box. 

The standard Ising model for  
(i) $d=2$ outside the closure of the coexistence line,   
(ii) $d=3$, $h\neq0$, and $T\ll T_c$, 
provides two examples where there is a diverging characteristic length 
as $T\to T_c$ and $h\to0$ respectively. In both cases restricted complete  
analyticity has been proven whereas complete analyticity in Dobrushin and  
Shlosman's sense has not yet been proven in the case (i) and  
actually disproven in the case (ii)  
\cite{[MO2], [MOS],[ScS]}. 
We finally note that an interesting direct proof of restricted  
complete analyticity, starting from a finite size condition similar  
to the one in \cite{[O],[OP]}, has been established in \cite{[M]}  
without the use of cluster expansion. 
 
The physical interest of 
the above discussion in connection with RG maps  
lies in the possibility of well defining a renormalization 
map for potentials close to their critical point. Actually, it 
is believed that even if the object system is critical it may happen that 
the constrained systems are in the one phase, weakly coupled regime, so that 
the renormalized potential is still well defined, 
see \cite{[HK],[BMO],[CiO]}. 
 
\medskip 
Let us go back to the discussion of the possible pathology of 
non--Gibbsianity. 
It frequently happens that the renormalized configuration $m$ inducing 
non--Gibbsianity, via violation of quasi locality, is very atypical 
with respect to renormalized measure $\nu^{(\ell)}$. It is then natural and 
physically relevant to introduce a weaker notion of Gibbsianity by 
requiring that conditional probabilities are well behaved only 
$\nu^{(\ell)}$ almost surely. 
More precisely weak Gibbsianity of $\nu^{(\ell)}$ means the following: 
there exists a ``good set" $\bar\cM^{(\ell)}\subset\cM^{(\ell)}$ with 
$\nu^{(\ell)}(\bar\cM^{(\ell)})=1$, such that for 
$m\in\bar\cM^{(\ell)}$ the conditional probabilities of the measure 
$\nu^{(\ell)}$ have the usual Gibbs form with respect to a  
potential $\{\Phi^{(\ell)}_I\}_{I\subset\subset\cL^{(\ell)}}$, 
$\Phi^{(\ell)}_I:\cM^{(\ell)}_I\to\bR$, 
satisfying the {\it pointwise} absolute summability: for each 
$i\in\cL^{(\ell)}$ and $m\in\bar\cM^{(\ell)}$ we have 
$\sum_{I\ni i}|\Phi^{(\ell)}_I(m_I)|<\infty$, 
but not the usual {\it uniform} absolute summability namely, 
$\sup_{i\in\cL^{(\ell)}} 
 \sum_{I\ni i} \sup_{m_I\in\cM^{(\ell)}_I} |\Phi^{(\ell)}_I(m_I)|<\infty$. 
The idea of looking at the weak Gibbs property goes 
back to Dobrushin \cite{[DRenkum]}.  
It has subsequently developed in \cite{[DS5]} and in many other papers, 
see for instance \cite{[MRSvM],[BKL]}. 
The main point of weak Gibbsianity is the construction of the set  
$\bar\cM^{(\ell)}$ of full $\nu^{(\ell)}$ measure. 
The key property is that for $m\in\bar\cM^{(\ell)}$  
the ``bad situations," giving 
rise to long range correlations in $\mu_m^{(\ell)}$, are very ``sparse" 
namely, larger and larger bad regions are farther and farther apart.  
 
We discuss the block averaging transformation:
it is known \cite{[EFS]} that the BAT renormalized measure for the 
Ising model at low temperature is 
non--Gibbsian because of violation of quasi locality induced by the configuration 
$m_i=0$ for all $i\in\cL^{(\ell)}$. It is clear that any constrained system and 
in particular the one corresponding to $m_i=0$ 
does not depend at all on the value of the magnetic field $h$. 
On the other hand if $h\neq0$ the configuration 
$m_i=0$ is very unlikely with respect to the renormalized 
measure $\nu^{(\ell)}$. Therefore, with high 
$\nu^{(\ell)}$--probability, the regions with 
$m_i=0$ are very sparse; however, with probability one  
there are arbitrarily large regions with bad magnetization $m_i=0$. 
Inside these regions the situation described by the constrained measure 
$\mu_m^{(\ell)}$ is close to a first order phase transition 
with long range order. 
This prevents the good, Gibbsian, behavior of the conditional probabilities 
of $\nu^{(\ell)}$ as well as the estimates of the renormalized potential, 
{\it uniform} in the renormalized conditioning configuration $m$. 
In contrast, it is reasonable to expect weak Gibbsianity of the  
renormalized system, indeed this is proven in \cite{[BCObat]}.  
 
\subsec{Disordered systems and Griffiths' singularity}{s:dsgs} 
\par\noindent 
The above scenario, leading to the replacement of the notion of 
Gibbsianity with the one of weak Gibbsianity, 
shares common features with disordered systems in presence of the so called 
Griffiths' singularity. 
 
Let us consider the case of high temperature 
Ising--like spin glasses. They are described by the following  
formal Hamiltonian 
\be{spinglass} 
H(\sigma)= 
-\sum_{x,y}J_{x,y}\s_x\s_y-h\sum_{x}\s_x, 
\end{equation} 
where $\s_x \in \{-1,+1\}$, $J_{x,y}$, for all $x,y\in\cL=\bZ^d$, 
are i.i.d.\ random variables, and $h\in\bR$.  
For the sake of simplicity we further specify the model by assuming 
$J_{x,y}=0$ for $|x-y|\neq1$ and 
$J_{x,y}\sim \cN (0,1)$, namely $J_{x,y}$ are Gaussian independent  
random variables with mean zero and variance one. 
We denote by $\mu^{(J)}$ the Gibbs measure corresponding to the Hamiltonian 
(\ref{spinglass}). 
The ``typical" (with respect to the disorder) interaction energy between 
neighboring spins is of order one so that for small inverse temperature 
$\b$ our random system is in the weak coupling regime. However, with 
probability one there are arbitrarily large regions where the 
random couplings $J_{x,y}$ take large positive values giving rise, inside 
these regions, to the behavior of a low temperature ferromagnetic 
Ising system with long range order. For a similar case, the one of 
a low temperature ferromagnetic diluted Ising system, it 
has been shown, see \cite{[Gr],[Suto]}, that the infinite 
volume specific free energy is infinitely differentiable 
but not analytical in $h$. 
This is a sort of infinite order phase transition called 
``Griffiths' singularity". 
 
For a high temperature spin glass with unbounded random couplings 
a similar behavior is expected. We also expect exponential decay of 
correlations with a non--random decay rate but with a 
random unbounded prefactor. 
More precisely, let us denote by $\Omega$ the collection of all  
$\ul{J}:=\{J_{x,y},\,x,y\in\cL,\,|x-y|=1\}$; 
in the above conditions we expect that there exist $m>0$ and  
a set $\bar\O\subset\Omega$ of full measure such that for each 
$\ul{J}\in\bar\O$, there exists a positive real 
$C(\ul{J})$ such that 
the spin correlations have the following bound 
\begin{equation} 
\label{decay} 
\big|\mu^{(J)}(\sigma_x;\sigma_y)\big|\leq C(\ul{J})\,\exp\{-m|x-y|\} 
\end{equation} 
 
There are several approaches to the analysis of disordered systems in the above 
regime, let us just quote the two papers 
\cite{[vDKP],[FI]}. In \cite{[vDKP]}
it is proven, via a very elegant method which does not use cluster expansion,
that (\ref{decay}) holds for some $C(\ul{J})$ having bounded expectation. 
Although the case of high temperature spin glass is covered,
there are some restrictions on the applicability of this method 
and the set of full measure where $C(\ul{J})$ is bounded is not 
explicitly constructed.
In \cite{[FI]}, that appeared several years before \cite{[vDKP]}, 
a more powerful and more widely applicable method 
is presented, involving a graded cluster expansion. 
The set of full measure $\bar\O$ is explicitly constructed
via a multi--scale analysis similar to the one 
introduced in \cite{[FS]} to study the Anderson localization. 
It emerges from the analysis developed in 
\cite{[FI]}, based on a hierarchy of ``scales of badness" 
that, with high probability, larger and larger bad regions are farther   
and farther apart and the largest scale of 
badness seen close to the origin is finite. 
The theory developed in \cite{[FI]} gives rise directly to 
estimates valid with probability one and requires very mild 
assumptions on the probability distribution  of random couplings. 
 
\subsec{A graded cluster expansion}{s:unif} 
\par\noindent 
To analyze disordered systems close 
to criticality and the weak Gibbs property of renormalized measures,
we need a graded cluster expansion based on a scale--adapted approach. 
The graded cluster expansion that is developed in the present paper is in
the same spirit as the one in \cite{[FI]}; we point out briefly the
main differences. (i) Whereas in \cite{[FI]} the first step 
(on the good region) is on scale one (e.g.\ high temperature/large
magnetic field), our first step uses instead a scale adapted expansion. This
allows to treat, in dimension two, Ising systems arbitrary close to
the coexistence line. (ii) The recursive classification of
the bad regions is somewhat different. In \cite{[FI]} three recursive
conditions are imposed: on the diameter, on the volume, and on the
inter--distance. We instead require only the diameter and inter--distance
conditions. The relative probability
estimates, proving that such a classification can be obtained with
probability one with respect to the disorder \cite{[BCObat]}, can
be easily derived in a general setup by a method analogous to that introduced 
in \cite{[vDK]}. 
(iii) In \cite{[FI]} the \emph{polymerization} of the spin
system is a preliminary step made on the whole lattice, the
relative cluster expansions are then carried out recursively;
we perform recursively both the polymerization and the cluster
expansion. 
(iv) We abstracted the relevant model independent assumptions for general finite 
state space, finite range spins systems. 
Accordingly, the graded cluster expansion is developed
with respect to a non--trivial reference measure. 

Let us describe a possible application of our graded cluster expansion 
to disordered systems. Consider the case of small random 
perturbations of a ferromagnetic system at a given temperature 
larger but arbitrarily close to the critical value.  
To be concrete consider a ferromagnetic  
two dimensional Ising system with zero magnetic field and coupling constants 
given by i.i.d.\ random variables for different bonds with distribution  
\begin{equation} 
\label{rp} 
J=\left\{ 
\begin{array}{ll} 
1 & \textrm{with probability } 1-p\\ 
J_0 & \textrm{with probability } p\\ 
\end{array} 
\right. 
\end{equation} 
Fix a temperature $T$ slightly larger than the critical value $T_c$ 
corresponding to a deterministic system with coupling constant one. 
To our knowledge the above described situation has never been  
studied in the literature. We expect the following result  
\cite{[BCOran]}:  
given $T>T_c$ there exists $p>0$ such that for 
any arbitrarily large $J_0\leq +\infty$ we can construct a convergent 
graded cluster expansion implying, in particular, the decay 
property (\ref{decay}). 
We also mention that, adapting the methods in \cite{[M]}, 
an effective finite size condition involving the quenched  
expectation of correlations can be obtained \cite{[M2]}. 
 
It is also clear that, when studying weak 
Gibbsianity of renormalized measures, 
in order to compute renormalized potentials as convergent series, 
we need a complete theory based on graded cluster expansion 
since the methods developed in \cite{[vDKP]}, which avoid  
the use of cluster expansion, are not sufficient for this purpose. 
On the other hand it is also clear that if we want to study 
weak Gibbsianity only assuming strong mixing, in particular for 
systems close to criticality and/or to study 
convergence properties of the iterates of RG maps, we need to 
consider a graded cluster expansion whose first scale is not one but, 
rather, depends on the parameters. 
In \cite{[BCObat]} we study the BAT transformation 
only assuming strong mixing of the object system. 
In the framework of a graded cluster expansion, with a sufficiently 
large minimal scale length, using scale--adapted expansion to 
treat the first step of the hierarchy, 
we establish the weak Gibbsianity of the renormalized measure. 
Moreover we show, in a suitable sense, convergence to a (trivial) 
fixed point of renormalized potential as the RG scale $\ell$ goes 
to infinity. 
Our results apply to the two--dimensional Ising model in the 
uniqueness region i.e., for $h\neq 0$ or $h =0,\, T > T_c$ and, 
in particular below $T_c$ where non--Gibbsianity has 
been proven in \cite{[EFS]}. 
At the moment, we are not able to cover the case $h=0$, $T<T_c$. 
 
In \cite{[MRSvM]} as well as in \cite{[BKL]} 
the authors establish weak Gibbsianity for measures arising from the 
application of general decimation transformations to a low 
temperature Ising or Pirogov--Sinai system.  
They have to analyze constrained systems on arbitrarily large scales but  
they have to choose a sufficiently low temperature and their minimal length  
is of order one. Therefore their methods work only very far below  
the critical point. 
In both papers the authors first fix the scale $\ell$ of RG transformation 
and then choose a sufficiently low temperature. 
In particular they both have to choose lower and lower 
temperatures starting from larger and larger RG scales. 
This behavior is not in agreement with the general RG philosophy. 
In \cite{[S1]} this anomaly is fixed, as it is shown 
that at a given sufficiently low temperature is 
enough to get weak Gibbsianity for all large 
enough scales. This approach is still based on a low 
temperature expansion and it is neither suited to approach 
the critical point nor to study convergence 
properties of the iterates of RG maps. 
 
\subsec{Synopsis}{s:syn} 
\par\noindent 
In the present paper we construct the graded cluster expansion 
that will be used to treat  
weak Gibbsianity for the block averaging transformation \cite{[BCObat]} 
and disordered systems in the Griffiths' phase \cite{[BCOran]}.  
Here there is no random disorder in the 
interactions; however, we suppose deterministically possible to analyze 
the bad interactions on suitable increasing scale lengths. 
We treat iteratively the regions of increasing badness and prove  
convergence of the expansion on the basis of 
suitable assumptions on the potential in the good region and 
sufficient ``sparseness" of bad regions. 
In \cite{[BCObat]} we prove that, 
with probability one with respect to the disorder or to the 
renormalized spin configuration, the situation 
is the one deterministically assumed in the present paper. 
 
The assumption that the system is weakly coupled on the  
complement of the bad region of the lattice namely, the good part, 
is here formalized by the  
following assumption. Let $\Delta$ be a finite subset of the good region 
and $Z_\Delta(\sigma)$ be the partition function in $\Delta$  
with boundary condition $\sigma$. We assume 
\begin{equation} 
\label{basic} 
\log Z_\Delta(\sigma)=\sum_{X\cap\Delta\neq\emptyset} V_{X,\Delta}(\sigma) 
\end{equation} 
where the effective potential $V_{X,\Delta}$ satisfies the following 
conditions:
\begin{itemize} 
\item[(a)] 
 given a finite subset $X\subset\bZ^d$ the functions  
 $V_{X,\Delta}$ are constant  
 w.r.t.\ $\Delta$ for the sets $\Delta$ with a fixed intersection with X;
\item[(b)] 
 the functions $V_{X,\Delta}$ have a suitable decay property w.r.t.\ $X$  
 uniformly in $\Delta$ and $\sigma$. 
\end{itemize} 
 
The expression (\ref{basic}) can be obtained via cluster expansion in  
the weak coupling (high temperature and/or small activity) region but it  
also holds in the more general situation of the scale--adapted cluster  
expansion discussed before.  
In the latter case it holds  
provided the volume $\Delta$ is a disjoint union of cubes whose side length  
equals the scale $L$ of the expansion.  
We also note that (\ref{basic}) implies  
one of the Dobrushin--Shlosman complete analyticity conditions  
\cite{[DS3]} namely, Condition IVa, see \cite{[BCOalb]}. 
In the applications we discussed above condition (\ref{basic})  
will be derived via a scale--adapted cluster expansion and therefore  
it will hold only for volumes $\Delta$ which are disjoint unions of cubes  
whose side length equals the scale $L$ of the expansion.  
However, by rescaling the lattice and redefining the single spin state  
space we reduce to the case in which (\ref{basic}) holds for any  
finite subset $\Delta$ of the good region, which is the basic assumption  
of the present paper.     
 
The main result concerns an expression, similar to (\ref{basic}), of the 
logarithm of the partition function on a generic finite subset of the whole  
lattice, possibly intersecting its bad region. 
Its characteristic feature, with 
respect to a usual low activity expansion, is that here 
polymers are geometrical objects living on arbitrarily large scale. 
This rules out the possibility to prove analyticity of the 
infinite volume free energy but would allow to prove infinite 
differentiability and exponential tree decay of semi--invariants 
\cite{[BCOalb]}  
with an unbounded prefactor as it is typical of Griffiths' phase. 
The proof is achieved by using condition (\ref{basic}) to integrate  
over the good region and by using the multi--scale geometry of the bad  
regions to recursively compute the effective interaction among them, i.e.\  
to recursively integrate over the bad spins.  
 
\medskip 
This paper is organized as follows: 
in Section~\ref{s:not} we introduce the model and state 
our results in Theorems~\ref{t:rispot} and \ref{t:decay}. 
The latter, whose proof is based on the cluster expansion of the  
logarithm of the partition function provided by the former,  
states the exponential decay 
of the semi--invariants for suitable local functions. 
The proof of Theorem~\ref{t:rispot} is achieved via the graded  
cluster expansion whose basic setup is introduced in  
Section~\ref{s:gce}; there we also state the related technical  
result in Theorem~\ref{t:mainceds}, whose proof is split into two parts: 
the algebraic structure of the computation is provided in 
Section~\ref{s:algebra}, while all convergence issues 
are discussed in Section~\ref{s:stime}. The proof of Theorem~\ref{t:mainceds} 
is completed at the end of Section~\ref{s:stime}. 
The Theorems~\ref{t:rispot} and \ref{t:decay} are finally proven in  
Section~\ref{s:prot}. 
 
\sezione{Notation and results}{s:not} 
\par\noindent 
In this Section we introduce 
the general framework, define precisely the model we shall consider, and state 
our main results. 
Given $a,b\in\bR$ we adopt the usual notation  
$a\wedge b:=\min\{a,b\}$ and $a\vee b:=\max\{a,b\}$.  
Given a set $A$ we let $|A|$ be its cardinality. 
 
\subsec{The lattice}{s:lat} 
\par\noindent 
For $x=(x_1,\cdots,x_d)\in\bR^d$ we let  
$|x|_1:=\sum_{k=1}^d |x_k|$ and 
$|x|_\infty:=\sup_{k=1,\cdots,d} |x_k|$. 
The spatial structure is modeled by the 
$d$--dimensional cubic lattice $\bL:=\Z^d$ in which we let $e_i, i=1, \dots,d$ 
be the coordinate unit vectors.  
We use $X^\complement:=\bL\setminus X$ to denote the complement of  
$X\subset\bL$. 
We use $X\subset\subset\bL$ to indicate that $|X|$ is finite. 
On $\bL$ we consider the distances 
$\disuno(x,y):=|x-y|_1$ and 
$\disinfinito(x,y):=|x-y|_\infty$ for $x,y\in\bL$. 
As usual for $X,Y\subset\bL$ we set  
$\disuno(X,Y):=\inf\{\disuno(x,y),\; x\in X,\; y\in Y\}$, 
$\disinfinito(X,Y):=\inf\{\disinfinito(x,y),\; x\in X,\; y\in Y\}$, 
$\diamuno(X):=\sup\{\disuno(x,x'),\; x,x'\in X\}$, and  
$\diaminfinito(X):=\sup\{\disinfinito(x,x'),\; x,x'\in X\}$. 
Moreover, given $r\ge1$ and $X\subset\bL$ we let 
$\partial_rX:=\{x\not\in X:\;\disinfinito(x,X)\le r\}$ be the 
$r$--{\em external boundary} of $X$ and 
$\clos{X}:=X\cup\partial_rX$ be the $r$--{\it closure} of $X$. 
 
For $x\in\bL$ and $m$ a positive real we let 
$Q_m(x):=\{y\in\bL:\,x_i\le y_i\le x_i+(m-1),\,\forall i=1,\dots,d\}$ 
the cube of side length $m$ with $x$ the site with smallest 
coordinates and 
$B_m(x):=\{y\in\bL:\,\disuno(y,x)\le m\}$ 
the ball of side length $2m+1$ centered at $x$. 
We shall denote $Q_m(0)$, resp.\ $B_m(0)$, simply by $Q_m$, resp.\ $B_m$. 
For each $X\subset\subset\bL$ we denote by 
$\env(X)\subset\subset\bL$ the smallest parallelepiped, 
with axes parallel to the coordinate directions, containing $X$. 
 
\subsec{The configuration space}{s:conf} 
\par\noindent 
For some applications, for instance the block averaging transformation, 
we have to deal with systems in which even the single spin space is not 
translationally invariant. We introduce the basic notation. 
We suppose given a collection of strictly 
positive integers $S_x$, $x\in\bL$, such that 
$S:=\sup_x S_x<+\infty$. 
The single spin configuration space is given by 
a finite set $\cS_x$, $|\cS_x|=S_x+1$, where $x\in\bL$. 
We identify $\cS_x$ with $\{0,1,\dots, S_x\}$ which we 
endow with the discrete topology. 
The configuration space in $\Lambda\subset\bL$ is 
$\cS_\Lambda\equiv\cS(\Lambda):=\otimes_{x\in \L}\cS_x$. 
Finally, the configuration space in $\bL$ is 
$\cS:=\otimes_{x\in\bL}\cS_x$, equipped with the product topology. 
Elements of $\cS$, called {\it configurations}, are denoted by 
$\sigma,\tau,\eta,\dots$. The integer 
$\sigma_x \equiv\sigma(x)$ is called value of the 
spin at the site $x$. 
For $\L\subset\bL$ and $\sigma\in\cS$ 
we denote by $\sigma_\Lambda$ the restriction 
of $\sigma$ to $\Lambda$. 
We denote by $\cF$ the Borel $\sigma$--algebra of $\cS$ and, 
for each $\Lambda\subset\bL$, we set 
$\cF_{\Lambda}:=\sigma\{\sigma_x,\, x\in\Lambda\}\subset\cF$. 
 
Let $m$ be a positive integer and 
$\L_1,\dots,\L_m\subset\bL$ be pairwise disjoint subsets of $\bL$; 
if $\sigma_i\in\cS_{\L_i}$, $i=1,\dots,m$, we denote by 
$\s_1\s_2\cdots\s_m$ 
the configuration in $\cS_{\L_1\cup\cdots\cup\L_m}$ given by 
$\s_1\s_2\cdots\s_m(x):=\sum_{i=1}^m\id_{\{x\in\L_i\}}\s_i(x)$, 
$x\in\Lambda_1\cup\cdots\cup\Lambda_m$. 
 
A function $f:\cS\rightarrow\bR$ is called a {\em local} function 
if and only if there exists $\Lambda\subset\subset\bL$ such that 
$f\in\cF_{\Lambda}$, namely $f$ is $\cF_{\Lambda}$--measurable for 
some finite set $\Lambda$. 
For $f$ a local function we shall denote by $\supp(f)$, 
the so called support of $f$, the smallest $\Lambda\subset\subset\bL$ 
such that $f\in\cF_{\Lambda}$. If $f\in\cF_\Lambda$ we shall 
sometimes misuse the notation by writing $f(\sigma_\Lambda)$ for 
$f(\sigma)$. 
For $f\in\cF$ we let $\|f\|_{\infty}:=\sup_{\sigma\in\cS}|f(\sigma)|$ 
be the sup norm of $f$. 
 
\subsec{The Gibbs state}{s:gibbs} 
\par\noindent 
A {\it potential} $U$ is a collection of local functions on $\cS$  
labelled by finite subsets of $\bL$, 
namely $U:=\{U_X\in\cF_X,\,X\subset\subset\bL\}$. 
We shall consider {\it finite range} potential i.e., 
there exists a real $R\ge0$, called {\em range},  
such that $U_X=0$ if $\diamuno(X)>R$. We, finally, introduce the norm 
$\|U\|:=\sup_{x\in\bL}\sum_{X\ni x}\|U_X\|_{\infty}$. 
In the sequel we shall always understand that the real $r$  
appearing in the definition of the boundary $\partial_rX$ of $X\subset\bL$  
is chosen so that $r\geq R$. 
We remark that we do not 
require the potential to be translationally invariant. 
 
Let $\Lambda\subset\subset\bL$, $\sigma\in\cS$ 
and consider the {\it Hamiltonian} 
\be{ham} 
H_{\L}(\sigma):=\sum_{X\cap\L\neq\es} U_X\(\sigma\) 
\end{equation} 
In this paper we shall consider only finite volume Gibbs measures 
defined as follows: 
let $\tau\in\cS$, 
the finite volume Gibbs measure $\mu_{\Lambda}^{\tau}$, 
with boundary condition $\tau$, is the probability measure on $\cS_{\Lambda}$ 
given by 
\be{ls-sone} 
\mu_{\L}^\t (\s):=\frac{1}{Z_{\L}(\t)} 
\exp\{+H_{\L}(\s_\L\tau_{\L^\complement})\} 
\end{equation} 
where $Z_\Lambda\in\cF_{\Lambda^\complement}$, 
called the {\it partition function}, is the normalization constant. 
Note that we have defined the Gibbs measure with a sign convention opposite to  
the usual one.  
 
\subsec{All's well $\dots$}{s:problem} 
\par\noindent 
Our aim is the computation of the partition function 
$Z_{\Lambda}(\tau)$ 
under the hypotheses that the Gibbs random field is weakly coupled only 
on a part 
of the lattice, that will be called {\it good} and 
denoted by $\bG_0$, under suitable geometric conditions on such $\bG_0$. 
We shall assume the system admits a convergent cluster expansion 
in $\bG_0$ with a 
suitable tree decay of the {\it effective} potential among the spins  
in $\bL\setminus\bG_0$ and resulting from the integration  
in $\bG_0$.  
 
Let 
$\bE:=\lg\{x,y\},\, x,y\in\bL:\, \disuno(x,y)=1 \rg$ 
the collection of {\em edges} in $\bL$. 
Note that, according to our 
definitions, the edges are parallel to the coordinate directions. 
We say that two edges $e,e'\in\bE$ 
are connected if and only if $e\cap e'\neq\emptyset$. 
A subset $(V,E)\su (\bL,\bE)$ is said to be connected iff 
for each pair $x,y \in V$, with $x\neq y$, 
there exists in $E$ a path of connected 
edges joining them.  
We agree that if $|V|=1$ then $(V,\emptyset)$ is connected. 
For $X\ssu\bL$ we then set 
\be{treedec0} 
\tree(X):=\inf\lg |E| \,, \  (V,E) \su (\bL,\bE) 
                \textrm{ is connected and } V \supset X 
\rg 
\end{equation} 
Note that $\tree(X)=0$ if $|X|=1$. 
We remark that for each $x,y\in\bL$ we have 
$\tree(\{x,y\})=\disuno(x,y)$. 
 
\bcon{t:ipotesi} 
Given $\bG_0\subset\bL$, we assume that for each 
$\Lambda\subset\subset\bL$ and $\sigma\in\cS$ 
we have the expansion 
\begin{equation} 
\label{ipo} 
Z_{\Lambda\cap\bG_0}(\sigma):= 
\sum_{\eta\in\cS(\Lambda\cap\bG_0)} 
\exp\Big\{ 
  H_{\Lambda\cap\bG_0}(\eta\sigma_{(\Lambda\cap\bG_0)^\complement})\Big\} 
= 
\exp\Big\{ 
\sum_{X\cap\Lambda\neq\emptyset} 
 V_{X,\Lambda}(\sigma)\Big\} 
\end{equation} 
for suitable local functions 
$V_{X,\Lambda}:\cS\rightarrow\bR$ satisfying the following properties 
\begin{enumerate} 
\item\label{i:cond2.1.1} 
given $\Lambda,\Lambda'\subset\subset\bL$ if 
$X\cap\Lambda=X\cap\Lambda'$ 
then $V_{X,\Lambda}=V_{X,\Lambda'}$; 
\item\label{i:cond2.1.2} 
$V_{X,\Lambda}\in\cF_{X\cap(\Lambda\cap\bG_0)^\complement}$; 
\item\label{i:cond2.1.3} 
$V_{X,\Lambda}=0$ if  
$X\cap\big(\clos{\Lambda}\big)^\complement\neq\emptyset$.  
\end{enumerate} 
Moreover, the {\it effective potential} 
$V_{\Lambda}:=\{V_{X,\Lambda},\, X\cap\Lambda\neq\emptyset\}$ can be bounded 
as follows: there are reals $\a>0$ and $A<\infty$ such that 
\be{ip2} 
\sup_{x\in\bL}\, 
 \sum_{X\ni x} e^{\alpha\tree(X)} 
  \sup_{\newatop{\Lambda\subset\subset\bL:} 
                {\Lambda\cap X\neq\emptyset}} 
   \|V_{X,\Lambda}\|_{\infty} 
\le A 
\end{equation} 
\econ 
We recall that our aim is to cluster expand 
$\log Z_{\Lambda}(\tau)$ 
with $\Lambda\subset\subset\bL$ and $\tau\in\cS$. Given 
$\Lambda\subset\subset\bL$, we first 
apply (\ref{ipo}) to the configuration 
$\sigma_\Lambda\tau_{\Lambda^\complement}$ 
and then we integrate on the variables  
$\sigma_{\Lambda\setminus(\Lambda\cap\bG_0)}$. 
 
\subsec{$\dots$ or sparse $\dots$}{s:multi} 
\par\noindent 
We shall not make any assumption on the behavior of the Gibbs field on the 
complement of the good part, namely the {\it bad} part of the lattice 
$\bB_0:=\bL\setminus\bG_0$, but 
we shall require that the bad sites are sparse enough. 
We start from the partition $\bL=\bG_0\cup\bB_0$ of the lattice 
in {\it good} and {\it bad} sites. Although such a sharp 
classification seems to be the reason for the never ending popularity 
of most American movies, it is not sufficient to our purposes. Let us 
forget about the good sites and look more closely at the bad 
ones. Some of them are not really bad, they are just bad guys far away 
from all the other bad sites (only close enough bad individuals form a 
dangerous gang). We are not really allowed to call such a behavior bad 
and we say they are {\it gentle} (more precisely 1--{\it 
gentle}). We next forget also about the 1--{\it gentle} sites and look 
at the remaining ones, which we call 1--{\it bad}. Even among them some 
are not so bad, after all. Maybe we have just a small group of bad 
guys very far away from all the 1--{\it bad} sites; those are called 
2--{\it gentle}. Proceeding in such a way we 
construct a multi--scale classification of the sites and we also suppose 
a happy ending: there are no $\infty$--{\it bad} guys. 
We formalize the above discussion in the following Definitions. 
 
\bdefi{t:seq} 
We say that two 
strictly increasing sequences $\G=\{\G_j\}_{j\ge 0}$ and 
$\g=\{\g_j\}_{j\ge 0}$ are {\em steep scales} iff they 
satisfy the following conditions: 
\begin{enumerate} 
\item\label{seq:<} 
$\Gamma_0=0$, $\gamma_0\ge0$, 
$\G_1\ge 2$, $\gamma_1>r\ge R$, and $\G_j < \g_j / 2 $ for any $j\ge 1$. 
\item\label{seq:con-dist} 
For $j\ge 0$ set 
${\displaystyle \vartheta_j:=\sum_{i=0}^j (\G_i+ \g_i)}$ and 
$\lambda:=\inf_{j\ge0}(\Gamma_{j+1}/\vartheta_j)$; then $\lambda>7$. 
\item\label{seq:covol} 
We have 
${\displaystyle \;\sum_{j=0}^\infty\frac{\G_j}{\g_j}\le\frac{1}{2}\;}$ 
where we understand $\Gamma_0/\gamma_0=0$ even in the case $\gamma_0=0$.  
\end{enumerate} 
\edefi 
 
\noi 
It is useful to remark that from items \ref{seq:con-dist} and  
\ref{seq:covol} above we get that 
\be{seq:D<g} 
\vartheta_j \le \g_j \bigg[1+\Big(1+\frac{1}{\lambda}\Big) 
              \sum_{i=1}^\infty\frac{\Gamma_i}{\gamma_i}\bigg]\le 2 \g_j, 
\qquad \textrm{for any } j\ge0  
\end{equation} 
Indeed, from item \ref{seq:con-dist} it follows  
$\gamma_j\le\Gamma_{j+1}/\lambda$; hence for $j\ge1$ we have  
\begin{equation*} 
\begin{array}{lcl} 
\vartheta_j & = &{\displaystyle 
         \g_j \bigg( \sum_{i=0}^j \frac{\G_i}{\g_j} + 1 + 
         \sum_{i=0}^{j-1} \frac{\g_i}{\g_j} \bigg) 
	 \vphantom{\bigg\}_{\big]}} } \\ 
    &\le&{\displaystyle 
           \g_j \bigg( 1 +\sum_{i=0}^j \frac{\G_i}{\g_i} 
	   +\frac{1}{\lambda}\sum_{i=0}^{j-1} \frac{\G_{i+1}}{\g_{i+1}} \bigg) 
	   \le \g_j \bigg(1+\frac{1}{2}+\frac{1}{2\lambda}\bigg)} 
\end{array} 
\end{equation*} 
 
\begin{rem} 
\label{t:sceltaseq} 
We note that the Conditions~\ref{seq:con-dist} and 
\ref{seq:covol} in the above definition, 
force a superexponential growth of the sequences $\Gamma$ and $\gamma$. 
It is easy to show that, given $\beta\ge9\vee(4/9)\log(8r)$, the sequences 
$\Gamma_0=\gamma_0:=0$,  
\begin{equation} 
\label{scletaseq} 
\Gamma_k:=e^{(\beta+1)(3/2)^k}\;\;\;\;\;\textrm{ and }\;\;\;\;\; 
\gamma_k:=\frac{1}{8}e^{\beta(3/2)^{k+1}} 
\;\;\;\;\;\textrm{ for } k\ge1 
\end{equation} 
provide an example of steep scales.  
\end{rem} 
 
\bdefi{t:gentle}  
We say that $\cG:=\{\cG_j\}_{j\ge 0}$, where each $\cG_j$ is a collection of  
finite subsets of $\bL$, is a {\em graded disintegration of} $\bL$ iff 
\begin{enumerate} 
\item\label{i:gd1} 
for each $g\in\bigcup_{j\ge0}\cG_j$  
there exists a unique $j\ge0$, which  
is called the {\em grade} of $g$, such that $g\in\cG_j$; 
\item\label{i:gd2} 
the collection $\bigcup_{j\ge0}\cG_j$  
of finite subsets of $\bL$ is a partition  
of the lattice $\bL$ namely, it is a collection of not empty pairwise disjoint  
finite subsets of $\bL$ such that  
\begin{equation} 
\label{part} 
\bigcup_{j\ge 0}\,\bigcup_{g\in\cG_j}g=\bL. 
\end{equation} 
\end{enumerate} 
\hfill\break 
Given $\bG_0\subset\bL$ and $\Gamma,\gamma$ steep scales, we say that  
a graded disintegration $\cG$ is a {\em gentle disintegration} of $\bL$  
with respect to $\bG_0,\Gamma,\gamma$ iff 
the following recursive conditions hold: 
\begin{enumerate} 
\setcounter{enumi}{2} 
\item 
\label{gent:null}  
$\cG_0=\big\{\{x\},\,x\in\bG_0\big\}$; 
\item 
\label{gent:diam}  
if $g\in\cG_j$ then $\diamuno(g) \le \G_j$ for any $j\ge 1$; 
\item 
\label{gent:dist} 
set $\bG_j:=\bigcup_{g\in\cG_j}g\subset\bL$, $\bB_0:=\bL\setminus\bG_0$ and  
$\bB_j:=\bB_{j-1}\setminus\bG_j$, then for  
any $g\in\cG_j$ we have $\disuno\(g,\bB_{j-1}\setminus g\)>\g_j$ for 
any $j\ge 1$; 
\item 
\label{gent:cas} 
for each $x\in\bL$ we have  
$k_x:=\sup\big\{j\ge 1:\,\exists g\in\cG_j\textrm{ such that } 
                \disinfinito(x,\env(g))\le \vartheta_j\big\}<\infty$,  
where we recall $\env(g)$ has been defined at the end of Section~\ref{s:lat}. 
\end{enumerate} 
Sites in $\bG_0$ (resp.\ $\bB_0$) are called {\em good} (resp.\ {\em bad}); 
similarly we call $j$--gentle (resp.\ $j$--bad)  
the sites in $\bG_j$ (resp.\ $\bB_j$). 
Elements of $\cG_j$, with $j\ge 1$, are called $j$--gentle atoms. 
Finally, we set  
$\cG_{\ge j}:= \bigcup_{i\ge j} \cG_i$. 
\edefi 
 
For $G\subset\cG_{\ge0}$ we define 
$\proj{G}:=\bigcup_{g\in G}g\subset\bL$; note that  
$\proj{\cG_j}=\bG_j$ and $\proj{\{g\}}=g$. 
Given the integers $j\ge0$, $s\ge 0$ and 
$G\subset\subset\cG_{\ge j}$, 
such that $G\cap\cG_j\neq\emptyset$, we 
define 
\be{esg0} 
\Es_s(G):=\big\{x\in\bL:\, 
                 \disinfinito(x,\env(\proj{G}))\le \vartheta_j+s\big\} 
\end{equation} 
Moreover for each $s\ge 0$ we set $\dEs_s(G):=\Es_s(G)\setminus\Es_{s-1}(G)$ 
where we understand $\Es_{-1}(G)=\emptyset$. 
 
\subsec{$\dots$ that ends well}{s:pres} 
\par\noindent 
As discussed before, our aim is to prove that, 
under Condition \ref{t:ipotesi} on the good part of the lattice and 
the sparseness condition formulated in Definitions~\ref{t:seq} and 
\ref{t:gentle} of the bad part of the lattice, the system 
admits a convergent cluster expansion.  
We set 
\begin{equation} 
\label{costa} 
a:=\frac{4d}{9(44)^{1/d}},\;\;\;\; 
q:=\frac{1}{2^5\,3^2}, 
\;\;\;\;\textrm{and}\;\;\;\; 
\varrho:=\Big\{\frac{1}{1-q}\bigg(1+\frac{1}{\alpha}\log A\bigg)\Big\}\vee0 
\end{equation} 
where we recall $\alpha$ and $A$ are the parameters in  
Condition~\ref{t:ipotesi}. 
 
\begin{teo} 
\label{t:rispot} 
Suppose Condition \ref{t:ipotesi} holds with $\alpha>0$ and $A<+\infty$ 
for some $\bG_0\subset\bL$. 
Assume also that 
for such $\bG_0$ there exist steep scales $\gamma,\Gamma$ 
and a gentle disintegration $\cG$ of $\bL$ with respect to 
$\bG_0,\Gamma,\gamma$ as in Definition~\ref{t:gentle}.  
Finally assume the scales $\Gamma,\gamma$ are such that 
\begin{enumerate} 
\item\label{i:dc3} 
we have 
$\Gamma_1>\max\{4(1+\log 3)/\alpha,\,(8d)^3/(2a)\}$; 
\item\label{cc2} 
we have 
$A (\Gamma_j+1)^de^{-\alpha\gamma_j/4}\le 1$ for any $j\ge 1$; 
\item\label{i:dc5} 
we have 
${\displaystyle 
  \sum_{j=1}^\infty\frac{8d}{a^{1/3}}\frac{1}{\gamma_j^{1/3}} 
  \le\frac{\alpha}{32}}$; 
\item\label{i:dc6} 
for each $j\ge 1$ we have 
${\displaystyle 
  \gamma_j\ge\bigg(\frac{j}{d}\bigg)^{3/2}a^{1/2}}$. 
\end{enumerate} 
Then, for each 
$X,\Lambda\subset\subset\bL$ there exist functions 
$\Psi_{X,\Lambda},\Phi_{X,\Lambda}\in\cF_{X\cap\Lambda^\complement}$ 
such that the following 
statements hold. 
\begin{enumerate} 
\item 
\label{p:tm1} 
For each $\Lambda\subset\subset\bL$  
we have the totally convergent expansion  
\begin{equation} 
\label{tm1} 
  \log Z_{\Lambda}(\tau)= 
  \sum_{X\cap\Lambda\neq\emptyset} 
     \left[\Psi_{X,\Lambda}(\tau)+\Phi_{X,\Lambda}(\tau)\right] 
\end{equation} 
\item 
\label{p:tm2} 
Let $\Lambda,\Lambda'\subset\subset\bL$, for each  
$X\subset\subset\bL$ such that 
$X\cap\Lambda=X\cap\Lambda'$ we have that 
$\Psi_{X,\Lambda}=\Psi_{X,\Lambda'}$ and 
$\Phi_{X,\Lambda}=\Phi_{X,\Lambda'}$. 
\item 
\label{p:tm3} 
Let $X,\Lambda\subset\subset\bL$,  
if $\diaminfinito(X)>\varrho$  
and there exists no $g\in\cG_{\ge 1}$ such that $\Es_0(g)=X$  
then $\Psi_{X,\Lambda}=0$.  
Moreover for each $x\in\bL$, recalling the integer $k_x$ has been introduced  
in item~\ref{gent:cas} of Definition~\ref{t:gentle},  
\begin{equation} 
\label{tm3} 
\sum_{X\ni x} 
 \sup_{\Lambda\subset\subset\bL} 
 \|\Psi_{X,\Lambda}\|_{\infty} 
\le 
A 
+ 
k_x(\Gamma_{k_x}+1+2\vartheta_{k_x})^{2d} 
   \big[\log S+\|U\|+k_x(1\vee A)(8^d+1)\big] 
\end{equation} 
\item 
\label{p:tm4} 
We have  
\begin{equation} 
\label{tm2} 
\sup_{x\in\bL} 
 \sum_{X\ni x}e^{q\alpha\diaminfinito(X)} 
 \sup_{\Lambda\subset\subset\bL} 
  \|\Phi_{X,\Lambda}\|_{\infty} 
\le  
e^{-\alpha}+e^{-q\alpha\gamma_1} 
             \Big(  
	          \frac{1+e^{-q\alpha/(2d)}}{1-e^{-q\alpha/(2d)}} 
             \Big)^d 
\end{equation} 
\end{enumerate} 
\end{teo} 
 
\noindent 
{\it Remark.}  
We note that for $\beta$ large enough, depending on $A$ and $\alpha$, 
the steep scales defined in  
Remark~\ref{t:sceltaseq} do satisfy items~\ref{i:dc3}--\ref{i:dc6}  
in the hypotheses of Theorem~\ref{t:rispot}.  
 
We next discuss the exponential decay of correlations which 
will be a simple consequence of the expansion in Theorem~\ref{t:rispot}.  
We stress that  
the decay of correlations cannot hold for all pairs of local functions; 
for instance, if their supports are contained in the same 
gentle atom, a possible long range order inside the atom itself 
could prevent such a decay. Our result essentially states 
the exponential decay of correlations except for such a case. 
In order to state this result we need few more definitions: 
let $\Lambda\subset\subset\bL$, $n\ge 2$ an integer,  
$f_1,\dots,f_n$ local functions such that  
$\supp(f_i)\subset\Lambda$ for $i=1,\dots,n$, 
$t_1,\dots,t_n\in\bR$, and $\tau\in\cS$; we define  
\be{pZ} 
Z_\Lambda\big(\tau;t_1,\dots,t_n\big) 
:= \mu_\Lambda^\tau\Big(\exp\Big\{\sum_{i=1}^n t_i f_i\Big\}\Big) 
\end{equation} 
The semi--invariant of $f_1,\dots,f_n$ with respect to the finite volume Gibbs 
measure $\mu_\Lambda^\tau$ is defined as 
\begin{equation} 
\label{2.1} 
\mu_\Lambda^\tau \big( f_1;\cdots;f_n\big) := 
\frac{\partial^n 
      \log Z_\Lambda\big(\tau;t_1,\dots,t_n)} 
{\partial t_1\cdots \partial t_n} 
\Bigg|_{t_1=\cdots=t_n=0} 
\end{equation} 
note that for $n=2$ we have  
$\mu_\Lambda^\tau \big( f_1;f_2\big) 
=\mu_\Lambda^\tau \big( f_1\, f_2\big) - \mu_\Lambda^\tau \big( f_1) 
\mu_\Lambda^\tau \big(f_2)$ namely, the covariance between $f_1$ and $f_2$. 
Let us denote, moreover, by $(\bV_n,\bE_n)$ the graph 
obtained from $(\bL,\bE)$ by contracting each $\supp(f_i)$, $i=1,\dots,n$, to a 
single point namely,  
$\bV_n:=[\bL\sm\bigcup_{i=1}^n\supp(f_i)]\cup\bigcup_{i=1}^n\{\supp(f_i)\}$,  
$\bE_n:=\{ \{v,v'\},\,v,v'\in\bV_n\,:\ \disuno(v,v')=1\}$, and set 
\begin{equation} 
\label{tif} 
T\big(f_1;\dots;f_n\big):= 
\inf\Big\{|E|,\, (V,E)\subset (\bV_n,\bE_n) 
\textrm{ connected and }V\supset\bigcup_{i=1}^n \{\supp(f_i)\} \Big\} 
\end{equation} 
 
\begin{teo} 
\label{t:decay} 
Suppose the hypotheses of Theorem~\ref{t:rispot} are satisfied. 
Let $n\in\bN$ and $f_1,\dots,f_n$ local functions such that the following 
conditions are satisfied: 
\begin{enumerate} 
\item 
\label{i:ris2} 
for each $i\neq j\in\{1,\dots,n\}$ we have  
$\disuno(\supp(f_i),\supp(f_j))>r\ge R$; 
\item 
\label{i:ris1} 
for each 
$i\neq j\in\{1,\dots,n\}$ there is no $g\in\cG_{\ge 1}$ such that 
$\Es_0(g)\cap\supp(f_i)\neq\emptyset$ and 
$\Es_0(g)\cap\supp(f_j)\neq\emptyset$. 
\end{enumerate} 
Then, there exist a real 
$M=M(A,\alpha,d,n;|\supp(f_1)|,\dots,|\supp(f_n)|)<+\infty$ such that 
\begin{equation} 
\label{adcs} 
\big|\mu_\Lambda^\tau(f_1;\dots;f_n)\big| 
   \le 
M 
\prod_{i=1}^n \mu_\Lambda^\tau(|f_i|)\: 
\exp\Big\{-\frac{q\alpha}{n-1}\: 
         T(f_1;\dots;f_n)\Big\} 
\end{equation} 
for any $\tau\in\cS$ and 
any $\Lambda\subset\subset\bL$ such that  
$\Lambda\supset\supp(f_i)$, $i=1,\dots,n$. 
\end{teo} 
 
\sezione{The graded cluster expansion}{s:gce} 
\par\noindent 
In this section we introduce our main technique, the graded cluster expansion, 
and state the related abstract results. 
It will be convenient to introduce the following notion. 
\bdefi{t:locmes} 
Given $X,V\subset\subset\bL$ and the family 
$F:=\{f_\Lambda:\cS\rightarrow\bR,\,\Lambda\subset\subset\bL\}$, we say 
that $F$ is $(X,V)$--{\em compatible} iff 
\begin{enumerate} 
\item 
\label{i:lm1} 
for each $\Lambda,\Lambda'\subset\subset\bL$ we have that 
$X\cap\Lambda=X\cap\Lambda'$ 
implies $f_\Lambda=f_{\Lambda'}$; 
\item 
\label{i:lm2} 
the function 
$f_\Lambda$ is $\cF_{(X\cap\Lambda^c)\cup V}$--measurable. 
\end{enumerate} 
\edefi 
\noindent 
In other words the family $\{f_\Lambda,\,\Lambda\subset\subset\bL\}$ is  
$(X,V)$--compatible if and only if $f_\Lambda$ does not change when  
$\Lambda$ is varied outside $X$ and it depends only on the configuration 
inside $V$ and the part of $X$ intersecting $\Lambda^c$. 
 
We suppose that $\cG$, as in Definition~\ref{t:gentle}, is 
a gentle disintegration of the lattice $\bL$ with respect to 
$\bG_0,\Gamma,\gamma$. 
We recall that a $j$--gentle atom $g\in\cG_j$ 
is a finite subset of $\bL$. 
If $G\ssu\cG_{\ge 1}$ by $|G|$ we always mean 
the cardinality of $G$ as a subset of $\cG_{\ge 1}$ 
i.e., the number of elements $g\in\cG_{\ge 1}$ in $G$.  
On the other 
hand, if $g\in\cG_{\ge 1}$ then $|g|$ denotes the cardinality of 
$g$ as a subset of $\bL$, but note that $|\{g\}|=1$. 
The building bricks of our polymers are finite subsets of 
$\cG_{\ge 1}$. From now on $s$ will always denote a positive integer. 
 
Given $X \ssu\bL$ we let $\xi(X)$ be the collection of 
the gentle atoms intersecting $X$ namely, 
\begin{equation} 
\label{csi} 
\xi(X):= 
\lg g\in\cG_{\ge 1}:\, g\cap X\neq\emptyset\rg 
\subset\subset\cG_{\ge 1} 
\end{equation} 
 
At scale $j$ the relevant notion of connectedness is the following. 
Given $G,G'\subset\cG_{\ge j}$ 
we say they are {\em $j$--connected}, 
and write $G\connj G'$, iff $G\cap G'\cap\cG_j\neq\emptyset$. 
A system $G_1,\dots,G_k$ with $G_h \su \cG_{\ge j}$ 
is said to be {\em $j$--connected} iff 
for each $h,h'\in\{1,\dots,k\}$ 
there exist $h_1,\dots,h_m\in\{1,\dots,k\}$ 
such that $G_h=G_{h_1}\connj G_{h_2}\connj\cdots\connj 
G_{h_m}=G_{h'}$. 
We are now ready to define the polymers at scale $j$ namely, we set  
\begin{equation} 
\label{polii} 
\begin{array}{l} 
\cR_j:=\bigcup_{k\ge 1}\big\{\{(G_1,s_1),\cdots,(G_k,s_k)\},\textrm{ where } 
             G_h\subset\cG_{\ge j},\ s_h\ge 0, \\ 
\phantom{\cR_j:=\bigcup_{k\ge 1}\big\{} 
  \textrm{for } h=1,\dots,k, 
  \textrm{ and the system } G_1,\dots,G_k \textrm{ is $j$--connected}\big\} 
\end{array} 
\end{equation} 
Elements of $\cR_j$ will be called {\em $j$--polymers}. 
Given a $j$--polymer $R=\{(G_1,s_1),\dots,(G_k,s_k)\}$ and 
$i\ge j$ we set 
$R\rest_{i}:=\bigcup_{h=1}^k G_h\cap\cG_i\subset\subset\cG_i$ 
and $R\rest_{\ge i} := 
\bigcup_{i'\ge i}R\rest_{i'}\subset\subset\cG_{\ge i}$. 
We also introduce the {\em support} of the polymer  
\begin{equation} 
\label{supol} 
\supp{R}:=\bigcup_{h=1}^k \Es_{s_h}(G_h)\subset\subset\bL 
\end{equation} 
We remark that a set $G\subset\cG_{\ge 1}$, with $|G|=n>1$,  
can be viewed as an $n$--body link, while $G=\{g\}$, with $g\in\cG_{\ge 1}$  
corresponds to one body. 
A pair $(G,s)$ has to be thought as a pair made of the link  
$G$ and the parallelepiped $\Es_s(G)$. The latter represents an  
``$s$--extended" support of the bond $G$.  
Thus the bricks of a polymer $R$ namely, the pairs $(G_h,s_h)$, can  
be viewed as the parallelepipeds $\Es_{s_h}(G_h)$ whose connectedness  
properties rely only upon the links $G_h$. The support of the  
polymer $R$, on the other hand, whose interest will become clear in the  
sequel, is defined as the union of the $s_h$--extended supports 
$\Es_{s_h}(G_h)$.  
 
Given two $j$--polymers $R,S\in\cR_j$ 
we say they are {\em $j$--compatible}, and 
write $R\, \compa_j \, S$, iff 
$R\rest_{j} \cap S\rest_{j} =\es$. Conversely we say that $R,S$ are 
{\em $j$--incompatible}, and write $R\,\inc_j \, S$ iff they are not 
{\em $j$--compatible}. 
We say that a collection $\ul{R}=\{R_1,\dots,R_k\}$, where 
$R_h\in\cR_j$, for $h=1,\dots,k$, of $j$--polymers forms a {\em 
cluster of $j$--polymers} 
iff it is not decomposable into two non empty 
subsets $\ul{R}= \ul{R}_1\cup\ul{R}_2$ such that every pair 
$R_1\in\ul{R}_1$, $R_2\in\ul{R}_2$ is $j$--compatible. 
We denote by $\ul{\cR}_j$ the collection of all 
the clusters of $j$--polymers. 
More precisely we have 
\begin{equation} 
\label{cRi} 
\begin{array}{l} 
\ul{\cR}_j:=\bigcup_{k\ge 1}\big\{\ul{R}= 
          \lg R_1,\dots,R_k\rg 
          \,,\; R_h\in\cR_j\, : \forall\; h,h'\in\{1,\dots,k\}\\ 
\phantom{\ul{\cR}_j:=\bigcup_{k\ge 1}\big\{} 
\exists\; 
h_1,\dots,h_m\in\{1,\dots,k\}\textrm{ such that } 
R_h=R_{h_1}\inc_j R_{h_2}\cdots\inc_j R_{h_m}=R_{h'}\big\}\\ 
\end{array} 
\end{equation} 
We remark that repetitions of the same $j$--polymer are allowed. 
We also define 
\be{cRiinc} 
\ul{\cR}_j^{\compa} 
:=\bigcup_{k\ge 1}\big\{\lg R_1,\dots,R_k\rg 
\,,\; R_h\in\cR_j \textrm{ such that } 
R_h \:\compa_j\: R_{h'},\, h\neq h'\big\} 
\end{equation} 
Given ${S}\in\cR_j$ and $\ul{R}\in\ul{\cR}_j$ we write 
$\ul{R}\, \inc_j \,{S}$ iff there exists $R\in \ul{R}$ such that 
${R}\, \inc_j\, {S}$. 
For $i\ge j$, $\ul{R}\in\ul{\cR}_j$ we set 
$\ul{R} \rest_{i} := \bigcup_{R\in\ul{R}} R \rest_{i}$, 
$\ul{R}\rest_{\ge i}:=\bigcup_{i'\ge i} \ul{R} \rest_{i'}$; 
we finally set $\supp\ul{R}:=\bigcup_{R\in\ul{R}}\supp R$. 
 
The setup introduced above is needed to develop the algebraic 
structure of the graded cluster expansion. In order to formulate 
the necessary recursive estimates, which quantify the decay of the 
effective interaction at scale $i$, we also need to take into account 
the couplings {\em below} scale $i$ and we need to introduce 
some more notation. 
Let $G=\{g_1,\dots,g_n\}\subset\subset\cG_{\ge 1}$, we set 
\begin{equation} 
\label{alb_iG} 
\cT(G):=\inf_{\newatop{x_m\in g_m}{m=1,\dots,n}} 
        \tree\(\{x_1,\dots,x_n\}\) 
\end{equation} 
 
We finally introduce some combinatorial factors as follows: 
for each $j\ge 1$, $k\ge 1$ and $\{R_1,\dots,R_k\}\in\ul{\cR}_j$ we set 
\be{inconoscibile} 
\f_T(R_1,\dots,R_k):=\frac{1}{k!} \sum_{f\in F(R_1,\dots,R_k)}(-1)^{ \# \; 
{\rm edges \;in} \;f} 
\end{equation} 
where $F(R_1,\dots,R_k)$ is the collection of connected subgraphs 
with vertex set $\{1,\dots,k\}$ 
of the graph with vertices $\{1,\dots,k\}$ and edges 
$\{h,h'\}$ corresponding to pairs $R_h, R_{h'}$ such that $R_h\inc_j R_{h'}$. 
We set the sum equal to zero if $F$ is empty and one if $k=1$. 
 
\bteo{t:mainceds} 
Suppose the hypotheses of Theorem~\ref{t:rispot} are satisfied. 
Then, there exist functions 
$Z_{g,\Lambda}^{(j)},\zeta_{\underline{R},\Lambda}:\cS\rightarrow\bR$, 
with $j\ge 1$, $g\in\cG_j$ and $\underline{R}\in\underline{\cR}_j$, such that 
$\underline{R}\rest_{\ge j+1}=\emptyset$, and 
$\Lambda\subset\subset\bL$, such that 
\begin{enumerate} 
\item 
\label{i:main1} 
for each $\tau\in\cS$ and $\Lambda\subset\subset\bL$ 
the free energy $\log Z_{\Lambda}(\tau)$ can be written 
as the absolutely convergent series 
\be{ZL} 
\log Z_{\L}(\tau)= 
  \sum_{ 
        \genfrac{}{}{0pt}{}{X\cap\Lambda\neq\emptyset:}{\xi(X)=\emptyset} 
       } 
        V_{X,\Lambda}(\tau) 
 +\sum_{j=1}^\k 
  \sum_{g\in\cG_j} 
    \log Z_{g,\L}^{(j)}(\tau) 
 +\sum_{j=1}^\k 
  \sum_{\stackrel{\ul{R}\in\ul{\cR}_j} 
           {\scriptscriptstyle \ul{R}\rest_{\ge j+1} = \es}} 
    \varphi_T \(\ul{R}\)\zeta_{\underline{R},\Lambda}(\tau) 
\end{equation} 
where 
$\k=\k(\Lambda)<\infty$ is the minimal integer $k$ such that 
$\clos{\Lambda}\cap\bigcup_{j\ge k+1}\bG_j=\emptyset$, so that 
$\Lambda$ admits 
the partition $\Lambda=\bigcup_{j=0}^\k\Lambda_j$, with $\L_j:=\L\cap\bG_j$; 
\item 
\label{i:main2} 
for each $j\ge 1$ and $g\in\cG_j$ 
the family 
$\{Z_{g,\Lambda}^{(j)},\,\Lambda\subset\subset\bL\}$ is 
$(\Es_0(g),\emptyset)$--compatible and each function 
$Z_{g,\Lambda}^{(j)}$ is identically 
equal to one whenever $\clos{g}\subset\Lambda^\complement$. 
For each $j\ge 1$ and 
$\underline{R}\in\underline{\cR}_j$, such that 
$\underline{R}\rest_{\ge j+1}=\emptyset$, 
the family 
$\{\zeta_{\ul{R},\Lambda},\,\Lambda\subset\subset\bL\}$ is 
$(\supp\ul{R},\emptyset)$--compatible 
and each function $\zeta_{\ul{R},\Lambda}$ is 
identically zero if there exists $R\in\ul{R}$, 
$(G,s)\in R$ and $g\in G$ such that 
$\clos{g}\subset\Lambda^\complement$; 
\item 
\label{i:main3} 
let 
$\varepsilon:=\exp\{-\alpha\gamma_1/8\}$; 
then  
\begin{equation} 
\label{eautopot} 
\sup_{\Lambda\subset\subset\bL} 
\|\log Z_{g,\Lambda}^{(j)}\|_{\infty} 
\le 
(\Gamma_j+1)^d\big[\|U\|+\log S\big]+j (1\vee A) (8^d+1)\Gamma_j^d 
\end{equation} 
and 
\begin{equation} 
\label{polbound} 
\sup_{\Lambda\subset\subset\bL} 
\|\z_{\underline{R},\Lambda}\|_{\infty} 
\le 
 \prod_{R\in\underline{R}}\prod_{(G,s)\in R} 
 \varepsilon^{|G|} 
 \exp\Big\{-\frac{\alpha}{16} 
      \Big[{\cT}(G) 
             +\frac{1}{2}\dis(\env(\proj{G}),\dEs_s(G))\Big]\Big\} 
\end{equation} 
\end{enumerate} 
\eteo 
 
\sezione{Algebra of the expansion}{s:algebra} 
\par\noindent 
In this Section we introduce the algebra of the graded cluster 
expansion without discussing any convergence issue, which will be dealt 
upon in Section \ref{s:stime}. We suppose the hypotheses of 
Theorem~\ref{t:mainceds} are satisfied. Moreover, for  
$\Lambda\subset\subset\bL$ we define the set  
\be{raffio} 
\Upsilon_{\Lambda}:= 
\{X\subset\subset\bL:\, X\cap\Lambda\neq\emptyset 
\textrm{ and } 
X\cap(\clos{\Lambda})^\complement=\emptyset 
\} 
\end{equation} 
by item~\ref{i:cond2.1.3} in Condition~\ref{t:ipotesi} we can rewrite  
(\ref{ipo}) as  
\begin{equation} 
\label{ipo2} 
Z_{\Lambda\cap\bG_0}\(\s\)= 
\exp\Big\{ 
\sum_{X\in\Upsilon_{\Lambda}} 
 V_{X,\Lambda}(\sigma)\Big\} 
\end{equation} 
Given $\Lambda\subset\subset\bL$, 
for $G\ssu\cG_{\ge1}$ and $s\ge0$ let us define 
\be{strangeX} 
\Upsilon_{\Lambda}(G,s):=\{X\in\Upsilon_\Lambda:\,  
\xi(X)=G,\,X\subset\Es_s(G),\,X\cap\dEs_s(G)\neq\emptyset  
\} 
\end{equation} 
In other words $\Upsilon_\Lambda(G,s)$ is the collection of the  
subsets $X$ of $\Es_s(G)$ intersecting $\Lambda$,  
all and only the atoms of the gentle disintegration in $G$, and the  
annulus $\dEs_s(G)$. It is easy to show that for each  
$\Lambda,\Lambda'\subset\subset\bL$ one has 
\begin{equation} 
\label{poi} 
\Lambda\cap\Es_s(G)=\Lambda'\cap\Es_s(G) 
\;\Longrightarrow\; 
\Upsilon_\Lambda(G,s)=\Upsilon_{\Lambda'}(G,s) 
\end{equation} 
Notice, finally, that if there exists $g\in G$ such that 
$\clos{g}\subset\L^\complement$ then $\Upsilon_{\Lambda}(G,s)=\emptyset$. 
 
Recalling that $V_{X,\Lambda}$ has been introduced in \eqref{ipo}, for 
$i\ge 1$, and $g\in\cG_{i}$ we define the following function  
\be{autopot} 
\Psi^{(i,0)}_{g,\L}:= 
 \sum_{X\in\Upsilon_{\Lambda}(g,0)} V_{X,\Lambda} 
\end{equation} 
Recalling Definition~\ref{t:locmes}, we have that  
(\ref{poi}) above and the items \ref{i:cond2.1.1} and \ref{i:cond2.1.2}  
of Condition~\ref{t:ipotesi} 
imply that the family 
$\{\Psi_{g,\Lambda}^{(i,0)},\,\Lambda\subset\subset\bL\}$ is  
$(\Es_0(g),g)$--compatible. 
Furthermore, if $\clos{g}\subset\L^\complement$ 
then $\Upsilon_{\Lambda}(g,0)=\emptyset$ implies $\Psi^{(i,0)}_{g,\L}=0$. 
We shall look at $\Psi^{(i,0)}_{g,\L}$ 
as the contribution to the self interaction of the $i$--atom $g$ 
due to the integration on scale $0$. It will not be expanded, but 
it will contribute to the reference (product) measure relative to  
the expansion at step $i$. 
 
For $i\ge 1$, $G\ssu\cG_{\ge i}$ such that 
$G\cap\cG_i\neq\emptyset$, and $s\ge 0$ we define 
\be{eq:veri-pot} 
{\F}^{(i,0)}_{G,s,\L}:= 
\begin{cases} 
{\displaystyle 
\sum_{X\in\Upsilon_{\Lambda}(G,s)} 
V_{X,\Lambda} 
}&\text{if} \quad  (|G|,s)\neq(1,0)\\ 
\phantom{merda} 0 &  \text{if} \quad  (|G|,s)=(1,0)\\ 
\end{cases} 
\end{equation} 
As before we get that the family  
$\{\Phi^{(i,0)}_{G,s,\Lambda},\,\Lambda\subset\subset\bL\}$ is 
$(\Es_s(G),\proj{G})$--compatible and 
that $\Phi^{(i,0)}_{G,s,\L}=0$ if there exists $g\in G$ 
such that $\clos{g}\subset\L^\complement$. 
We shall look at $\Phi^{(i,0)}_{G,s,\L}$ 
as the effective interaction at scale $i$ due 
to the integration on scale $0$; it will be expanded at step $i$. 
 
By using definitions (\ref{autopot}) and (\ref{eq:veri-pot}) 
we have 
\begin{equation} 
\label{eq:tilde} 
\sum_{X\in\Upsilon_{\Lambda}} 
V_{X,\Lambda} 
=\sum_{\newatop{X\in\Upsilon_{\Lambda}:}{\xi(X)=\emptyset}} 
      V_{X,\Lambda}+\sum_{i\ge 1}\sum_{g\in\cG_i} 
      \Ps^{(i,0)}_{g,\L} 
+\sum_{i\ge 1} 
 \sum_{\newatop{G\subset\subset\cG_{\ge i}}{G\cap\cG_i\neq\emptyset}} 
 \sum_{s\ge 0} 
 \F^{(i,0)}_{G,s,\L} 
\end{equation} 
Note that if $\xi(X)=\emptyset$ and $X\subset\Lambda$ then 
$V_{X,\Lambda}\in\cF_\emptyset$, namely the function 
$V_{X,\Lambda}$ is constant. 
Moreover, 
since $\L\ssu\bL$, all but a finite number of terms 
on the r.h.s.\ of \eqref{eq:tilde} are vanishing. 
 
To simplify the notation 
for each $g\in\cG_{\ge 1}$ we define the bare self--interaction 
inside $g$ as  
\begin{equation} 
\label{Ug} 
U_{g,\L}:= 
\sum_{\newatop{X\subset\subset\bL:\, 
      X\cap\L\neq\emptyset}{X\cap\L\subset g\cap\L}} 
U_X 
\end{equation} 
and remark that, since the potential $U$ has range $R\le r$, we have that 
the family  
$\{U_{g,\L},\,\Lambda\subset\subset\bL\}$  
is $(\clos{g},g)$--compatible; furthermore, 
$U_{g,\L}=0$ if $g\cap\L=\emptyset$. 
 
Note that for $g,h\in\cG_{\ge 1}$, $g\neq h$, we have 
$\disuno(g,h)>\gamma_1>r\ge R$. 
Recalling that the integer $\k$ has been defined in Theorem~\ref{t:mainceds} 
we have $\cS_\L=\bigotimes_{j=0}^\k\cS(\L_j)$, 
where we recall in item~\ref{i:main1} of Theorem~\ref{t:mainceds}  
we have defined $\Lambda_j:=\Lambda\cap\bG_j$ for $j\ge0$.  
For $i\ge0$ we also set  
$$ 
\Lambda_{\ge i}:=\bigcup_{j\ge i}\Lambda_j 
\;\;\;\;\textrm{ and }\;\;\;\; 
\Delta_i:=\Lambda_{\ge i+1}\cup\Lambda^\complement 
$$ 
Then, given $\tau\in\cS$, 
recalling the abuse of notation mentioned at the end of  
Section~\ref{s:conf}, we have  
\begin{eqnarray} 
Z_{\Lambda}(\tau)&:=& 
\sum_{\eta\in\cS:\,\eta_{\Lambda^\complement}=\tau} 
 e^{H_\Lambda(\eta)}= 
\sum_{\newatop{\eta^\k\in\cS(\Delta_{\k-1})} 
              {\eta^\k_{\Delta_{\k}}=\tau}} 
\cdots 
\sum_{\newatop{\eta^1\in\cS({\Delta_0})} 
              {\eta^1_{\Delta_{1}}=\eta^2}} 
\sum_{\newatop{\eta^0\in\cS} 
              {\eta^0_{\Delta_{0}}=\eta^1}} 
 e^{H_\Lambda(\eta_0)} 
\nn\\ 
&=& 
\sum_{\newatop{\eta^\k\in\cS(\Delta_{\k-1})} 
              {\eta^\k_{\Delta_{\k}}=\tau}} 
\prod_{g\in\cG_\k}e^{U_{g,\L}(\eta^\k)} 
\cdots 
\sum_{\newatop{\eta^1\in\cS({\Delta_0})} 
              {\eta^1_{\Delta_{1}}=\eta^2}} 
\prod_{g\in\cG_1}e^{U_{g,\L}(\eta^1)} 
\sum_{\newatop{\eta^0\in\cS} 
              {\eta^0_{\Delta_{0}}=\eta^1}} 
 e^{H_{\Lambda_0}(\eta^0)} 
\end{eqnarray} 
Now, by using equations \eqref{ipo2} and \eqref{eq:tilde} we get 
\begin{equation} 
\label{zzzz} 
\begin{array}{rcl} 
{\displaystyle 
Z_{\Lambda}(\tau)} 
&=& 
{\displaystyle 
\exp\Bigg\{\sum_{\newatop{X\in\Upsilon_\Lambda:} 
                     {\xi(X)=\emptyset}} 
             V_{X,\Lambda}(\tau)\Bigg\}}\\ 
&& 
{\displaystyle 
\times 
\sum_{\newatop{\eta^\k\in\cS(\Delta_{\k-1})} 
              {\eta^\k_{\Delta_{\k}}=\tau}} 
\prod_{g\in\cG_\k} 
 e^{U_{g,\L}(\eta^\k)+\Psi^{(\k,0)}_{g,\L}(\eta^\k)} 
 \,\cdot\, 
 \exp\Bigg\{\sum_{\newatop{G\subset\subset\cG_{\ge \k}} 
                          {G\cap\cG_\k\neq\emptyset}} 
            \sum_{s\ge 0}\F^{(\k,0)}_{G,s,\L}(\eta^\k)\Bigg\} }\\ 
&& 
{\displaystyle 
\times\cdots\times 
\sum_{\newatop{\eta^1\in\cS(\Delta_{0})} 
              {\eta^1_{\Delta_{1}}=\eta^2}} 
\prod_{g\in\cG_1} 
 e^{U_{g,\L}(\eta^1)+\Psi^{(1,0)}_{g,\L}(\eta^1)} 
 \,\cdot\, 
 \exp\Bigg\{\sum_{\newatop{G\subset\subset\cG_{\ge 1}} 
                          {G\cap\cG_1\neq\emptyset}} 
            \sum_{s\ge 0}\F^{(1,0)}_{G,s,\L}(\eta^1)\Bigg\}} 
\end{array} 
\end{equation} 
 
We next define by recursion on $j=0,\dots,\k$ some functions 
$\F^{(i,j)}$ and $\Ps^{(i,j)}$, $0\le j < i \le \k$. As in the case 
$j=0$ we look at $\F^{(i,j)}$ as the effective interaction at scale 
$i$ due to the integration on scale $j<i$; on the other hand we look 
at $\Ps^{(i,j)}$ as the effective self--interaction at scale $i$ due to the 
integration on scale $j<i$. 
 
As recursive hypotheses we assume that  
we have already defined the families of functions  
$\{\Ps^{(i,m)}_{g,\L},\,\Lambda\subset\subset\bL\}$, which is   
$(\Es_0(g),g)$--compatible, and   
$\{\F^{(i,m)}_{G,s,\L},\,\Lambda\subset\subset\bL\}$, which is  
$(\Es_s(G),\proj{G})$--compatible,  
for any $m=0,\dots,j-1$, any $i=m+1,\dots,\k$, any $g\in\cG_i$, 
any $G\subset\subset\cG_{\ge i}$, such that 
$G\cap\cG_i\neq\emptyset$, and any $s\ge 0$.  
Moreover we assume 
$\Psi^{(i,m)}_{g,\Lambda}=0$ if $\clos{g}\subset\Lambda^\complement$ 
and  
$\Phi^{(i,m)}_{G,s,\Lambda}=0$ if $(|G|,s)=(1,0)$ or there exists 
$g\in G$ such that $\clos{g}\subset\Lambda^\complement$. 
We next define, by 
integrating on the scale $j$, the potentials 
$\Ps^{(i,j)}_{g,\L}$ and 
$\F^{(i,j)}_{G,s,\L}$ for $i=j+1,\dots,\k$, 
any $g\in\cG_i$, any $G\subset\subset\cG_{\ge i}$, such that 
$G\cap\cG_i\neq\emptyset$, and $s\ge 0$, and show that they satisfy  
the compatibility properties stated above.  
 
By the recursive assumptions and the properties of $U_{g,\Lambda}$, 
for each $g\in\cG_j$ the family of functions  
$\{U_{g,\L}+\sum_{m=0}^{j-1}\Psi^{(j,m)}_{g,\L},\,\Lambda\subset\subset\bL\}$ 
is $(\Es_0(g),g)$--compatible and a function of the family  
is identically zero if $\clos{g}\subset\Lambda^\complement$. Therefore, 
for $\eta^{j+1}\in\cS_{\Delta_j}$ we can set 
\be{ZG} 
Z_{g,\L}^{(j)}(\eta^{j+1}):= 
\sum_{\newatop{\eta^j\in\cS(g\cup\Lambda_{\ge j+1}\cup\Lambda^\complement)} 
              {\eta^j_{\Delta_j}=\eta^{j+1}}} 
\exp\lg 
U_{g,\L}(\eta^j) 
+\sum_{m=0}^{j-1}\Psi^{(j,m)}_{g,\L}(\eta^j) 
\rg 
\end{equation} 
We note that the family  
$\{Z_{g,\Lambda}^{(j)},\,\Lambda\subset\subset\bL\}$  
is $(\Es_0(g),\emptyset)$--compatible and a function of the family  
is identically 
equal to one if $\clos{g}\subset\Lambda^\complement$. 
For each $\eta^{j+1}\in\cS_{\Delta_j}$ we can define 
a probability measure $\nu_{g,\Lambda,\eta^{j+1}}^{(j)}$ on 
$\cS_g$ by setting, for each $\sigma\in\cS_g$, 
\be{nuG} 
\nu^{(j)}_{g,\L,\eta^{j+1}}(\sigma) 
:= 
\delta_{\eta^{j+1}}(\sigma_{g\cap\Lambda^\complement}) 
\frac{1}{Z_{g,\L}^{(j)}(\eta^{j+1})} 
\exp\lg 
U_{g,\L}(\sigma\eta^{j+1}) 
+\sum_{m=0}^{j-1}\Psi^{(j,m)}_{g,\L}(\sigma\eta^{j+1}) 
\rg 
\end{equation} 
For each $\sigma\in\cS_g$ the family  
$\{\eta^{j+1}\mapsto\nu_{g,\Lambda,\eta^{j+1}}(\sigma),\, 
                                \Lambda\subset\subset\bL\}$  
is $(\Es_0(g),\emptyset)$--compatible; moreover, 
$\nu_{g,\Lambda,\eta^{j+1}}=\delta_{\eta^{j+1}}$ 
if $g\subset\Lambda^\complement$. 
 
Given $G\su\su\cG_{\ge j}$ such that $G\cap\cG_j\neq\es$, 
and $s\ge 0$ we set 
\be{intj} 
\F^{(j)}_{G,s,\L}:= 
\sum_{m=0}^{j-1} \F^{(j,m)}_{G,s,\L} 
\end{equation} 
which is the (cumulated) effective interaction at scale $j$. 
By the recursive hypotheses we have that the family  
$\{\F^{(j)}_{G,s,\L},\,\Lambda\subset\subset\bL\}$  
is $(\Es_s(G),\proj{G})$--compatible; 
moreover $\F^{(j)}_{G,s,\L}$ is identically zero 
if $(|G|,s)=(1,0)$ or there exists $g\in G$ such that 
$\clos{g}\subset\Lambda^\complement$. 
 
Let $\eta^{j+1}\in\cS(\Delta_j)$ and  
$R=\{(G_1,s_1),\dots(G_k,s_k)\}\in\cR_j$; we define its activity 
$\zeta_{R,\L}(\eta^{j+1})$ as 
\be{acRj} 
\zeta_{R,\L}(\eta^{j+1}) 
:= 
\sum_{\newatop{\eta^j\in 
          \cS(\proj{R\rest_j}\cup\Lambda_{\ge j+1}\cup\Lambda^\complement)} 
              {\eta^j_{\Delta_j}=\eta^{j+1}}} 
\; 
\prod_{g\in R\rest_j} 
\nu^{(j)}_{g,\L,\eta^{j+1}}(\eta^j_g) 
\prod_{h=1}^k \[ \exp\lg 
\F^{(j)}_{G_h,s_h,\L} (\eta^j) 
\rg -1\] 
\end{equation} 
It follows that 
$\{\zeta_{R,\Lambda},\,\Lambda\subset\subset\bL\}$  
is $(\supp R,\proj{R\rest_{\ge j+1}})$--compatible and an element of  
the family is identically zero if there exists $(G,s)\in R$ and $g\in G$  
such that $\clos{g}\subset\Lambda^\complement$.  
For $\underline{R}\in\underline{\cR}_j$, we set 
\be{accrj} 
\zeta_{\ul{R},\L}(\eta^{j+1}) 
:=\prod_{R\in\ul{R}}\zeta_{R,\L}(\eta^{j+1}) 
\end{equation} 
it follows that 
$\{\zeta_{\ul{R},\Lambda},\,\Lambda\subset\subset\bL\}$ is 
$(\supp\ul{R},\proj{\ul{R}\rest_{\ge j+1}})$--compatible 
and an element of the family is identically zero if there exists  
$R\in\ul{R}$, $(G,s)\in R$, and $g\in G$ such that 
$\clos{g}\subset\Lambda^\complement$. 
 
By standard polymerization and cluster expansion, 
under suitable ``small activity" conditions that will be specified later on, 
see item \ref{conv:7} in Lemma~\ref{t:condsupp} below, we 
have, see e.g.\ \cite{[GJ]}, 
\bea{polce} 
&& 
\sum_{\newatop{\eta^j\in\cS(\Delta_{j-1})} 
              {\eta^j_{\Delta_j}=\eta^{j+1}}} 
\prod_{g\in\cG_j} 
\nu^{(j)}_{g,\L,\eta^{j+1}}(\eta^j_g) 
\exp\lg 
\sum_{\stackrel{G\ssu \cG_j} 
               {\scriptscriptstyle G\cap\cG_j\neq\es}} 
\sum_{s\ge 0} 
\F^{(j)}_{G,s,\L}(\eta^j) 
\rg 
\nonumber\\ 
&& 
\qquad\qquad\qquad 
=1+\sum_{\ul{R}\in \ul{\cR}_j^{\compa}} 
 \zeta_{\ul{R},\L}(\eta^{j+1}) 
=\exp\lg 
\sum_{\ul{R}\in \ul{\cR}_j} 
\f_T \( \ul{R} \) 
 \zeta_{\ul{R},\L}(\eta^{j+1}) 
\rg 
\end{eqnarray} 
with $\varphi_T$ defined in (\ref{inconoscibile}). 
 
We are now ready to define the interactions due to the integration on the 
scale $j$.  
Let $G\su\subset\cG_{\ge j+1}$ and $s\ge 0$, we define 
\be{cjp->Ys} 
\ul{\cR}_j(G,s):= 
\lg \ul{R}\in \ul{\cR}_j: \ 
\ul{R}\rest_{\ge j+1} = G\, , \ \supp\ul{R}\su \Es_s(G) ,\ 
\supp\ul{R} \cap \dEs_s(G) \neq\es 
\rg 
\end{equation} 
For $g\in\cG_i$, $i>j$, we let 
\be{selfintij} 
\Psi^{(i,j)}_{g,\L} 
:=\sum_{\ul{R}\in\ul{\cR}_j(g,0)} 
  \f_T\(\ul{R}\)\,\zeta_{\ul{R},\L} 
\end{equation} 
It is easy to check that 
$\{\Psi^{(i,j)}_{g,\Lambda},\,\Lambda\subset\subset\bL\}$  
is $(\Es_0(g),g)$--compatible and an element of the family is identically 
zero if $\clos{g}\subset\Lambda^\complement$; so we 
met the first recursive condition. 
The effective interaction at scale $i>j$ due 
to the integration on scale 
$j$ is defined as follows; for $G\ssu\cG_{\ge i}$, 
$G\cap\cG_i\neq\es$ and $s\ge 0$ we set 
\be{effiintij} 
\F^{(i,j)}_{G,s,\L} 
:= 
\begin{cases} 
{\displaystyle 
\sum_{\ul{R}\in\ul{\cR}_j(G,s)}\f_T\(\ul{R}\)\,\zeta_{\ul{R},\L} 
}&\text{if} \quad  (|G|,s)\neq(1,0)\\ 
\phantom{merda} 0 &  \text{if} \quad  (|G|,s)=(1,0)\\ 
\end{cases} 
\end{equation} 
As before 
$\{\Phi^{(i,j)}_{G,s,\Lambda},\,\Lambda\subset\subset\bL\}$  
is $(\Es_s(G),\proj{G})$--compatible and 
an element of the family is  
identically zero if there exists $g\in G$ such that 
$\clos{g}\subset\Lambda^\complement$; so we also 
met the second recursive condition. 
 
By noticing that 
\be{riarr} 
\sum_{\ul{R}\in \ul{\cR}_j} \f_T(\ul{R}) \zeta_{\ul{R},\L} 
= 
\sum_{\stackrel{\ul{R}\in \ul{\cR}_j} 
               {\scriptscriptstyle \ul{R}\rest_{\ge j+1}=\emptyset}} 
                   \f_T(\ul{R}) 
                      \zeta_{\ul{R},\L} 
+ 
\sum_{i=j+1}^\k 
\bigg\{ 
\sum_{g\in\cG_i} 
\Psi^{(i,j)}_{g,\L} 
+ 
\sum_{\stackrel{G\ssu \cG_{\ge i}} 
               {\scriptscriptstyle G\cap\cG_i\neq\es}} 
\sum_{s\ge 0} 
\F^{(i,j)}_{G,s,\L} 
\bigg\} 
\end{equation} 
and using recursively \eqref{polce} in \eqref{zzzz}, 
it is easy to check that, provided 
all the series converges absolutely, we have got the expansion 
(\ref{ZL}). 
 
\sezione{Convergence of the graded cluster expansion}{s:stime} 
\par\noindent 
In this section we prove the convergence of the cluster expansion 
introduced in Section \ref{s:algebra} above. 
 
\subsec{Geometric bounds}{s:gra} 
\par\noindent 
In this section we collect bounds which hold in our 
geometry of wide separated {\em gentle atoms}. 
For the reader convenience we restate \cite[Lemma~3.4]{[BCOalb]} 
in the present context.  
 
\blem{t:infpi} 
Let $k$ be a positive integer and 
$\Pi_0(k)$ be the set of permutations $\pi$ 
of $\,\{0,1,\dots,k\}$ such that $\pi(0)=0$. 
Let $X=\{x_0,x_1,\dots,x_k\}\subset\bL$ and $\tree(X)$ as in 
(\ref{treedec0}); then 
\begin{equation} 
\label{infpi0} 
\tree(X)\ge 
  \frac{1}{2} 
  \inf_{\pi\in\Pi_0(k)} 
  \sum_{l=1}^{k} \disuno\left(x_{\pi(l-1)},x_{\pi(l)}\right) 
\end{equation} 
\elem 
 
\setlength{\unitlength}{1.3pt} 
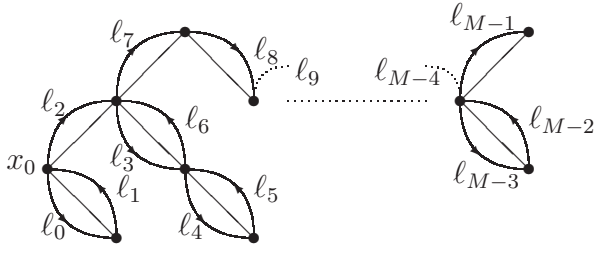
\begin{figure} 
\begin{picture}(200,120)(-120,-50) 
\thinlines 
\put(-20,20){\circle*{3}} 
\put(0,40){\circle*{3}} 
\put(0,0){\circle*{3}} 
\put(20,60){\circle*{3}} 
\put(20,20){\circle*{3}} 
\put(40,40){\circle*{3}} 
\put(40,0){\circle*{3}} 
\put(100,40){\circle*{3}} 
\put(120,60){\circle*{3}} 
\put(120,20){\circle*{3}} 
\put(-20,20){\line(1,1){20}} 
\put(-20,20){\line(1,-1){20}} 
\put(0,40){\line(1,1){20}} 
\put(0,40){\line(1,-1){20}} 
\put(20,20){\line(1,-1){20}} 
\put(20,60){\line(1,-1){20}} 
\put(100,40){\line(1,1){20}} 
\put(100,40){\line(1,-1){20}} 
\thinlines 
\qbezier(-20,20)(-20,0)(0,0)       \put(-15,5){\vector(1,-1){1}} 
\qbezier(-20,20)(0,20)(0,0)        \put(-5,15){\vector(-1,1){1}} 
\qbezier(-20,20)(-20,40)(0,40)     \put(-15,35){\vector(1,1){1}} 
\qbezier(0,40)(0,20)(20,20)        \put(5,25){\vector(1,-1){1}} 
\qbezier(20,20)(20,0)(40,0)        \put(25,5){\vector(1,-1){1}} 
\qbezier(40,0)(40,20)(20,20)       \put(35,15){\vector(-1,1){1}} 
\qbezier(20,20)(20,40)(0,40)       \put(15,35){\vector(-1,1){1}} 
\qbezier(0,40)(0,60)(20,60)        \put(5,55){\vector(1,1){1}} 
\qbezier(20,60)(40,60)(40,40)      \put(35,55){\vector(1,-1){1}} 
\qbezier[10](40,40)(40,50)(50,50)   
\qbezier[20](50,40)(70,40)(90,40)   
\qbezier[10](90,50)(100,50)(100,40) 
\qbezier(100,40)(100,20)(120,20)   \put(105,25){\vector(1,-1){1}} 
\qbezier(120,20)(120,40)(100,40)   \put(115,35){\vector(-1,1){1}} 
\qbezier(100,40)(100,60)(120,60)   \put(105,55){\vector(1,1){1}} 
\put(-32,20){$x_0$} 
\put(-22,1){$\ell_0$} 
\put(0,12){$\ell_1$} 
\put(-22,37){$\ell_2$} 
\put(-2,21){$\ell_3$} 
\put(18,1){$\ell_4$} 
\put(40,12){$\ell_5$} 
\put(20,32){$\ell_6$} 
\put(-2,57){$\ell_7$} 
\put(40,52){$\ell_8$} 
\put(52,46){$\ell_9$} 
\put(75,46){$\ell_{M-4}$} 
\put(98,16){$\ell_{M-3}$} 
\put(120,32){$\ell_{M-2}$} 
\put(97,62){$\ell_{M-1}$} 
\end{picture} 
\vskip -1.5 cm 
\caption{The path $\ell=\{\ell_0,\dots,\ell_{M-1}\}$ introduced 
in the proof of Lemma~\ref{t:infpi}.The solid circles represent  
the points $\{x_0,x_1,\dots,x_k\}$.} 
\label{f:alb} 
\end{figure} 
 
\medskip 
\noi{\it Proof.}\ 
It is easy to show that 
the infimum in \eqref{treedec0} is attained (not 
necessary uniquely) for a graph $T_X=(V_X,E_X)\subset(\bL,\bE)$ 
which is a tree, i.e.\ a connected and loop--free graph. 
The Lemma follows from the bound 
\begin{equation} 
\label{alb} 
|E_X|\ge 
  \frac{1}{2} 
  \inf_{\pi\in\Pi_0(k)} 
  \sum_{l=1}^{k} \disuno\left(x_{\pi(l-1)},x_{\pi(l)}\right) 
\end{equation} 
which is proven as follows. By induction on the number of edges in $T_X$ 
it is easy to prove, see Fig.~\ref{f:alb}, 
that there exists a path 
$(\ell_0,\dots,\ell_{M-1})$, with $\ell_m\in E_X$ for all $m=0,\dots,M-1$, 
satisfying the following properties: 
$\ell_{m-1}\cap\ell_m\neq\emptyset$ for all $m=1,\dots,M-1$, 
$x_0\in\ell_0$, for each $v\in V_X$ there exists 
$m\in\{0,\dots,M-1\}$ such that 
$\ell_m\ni v$, and each $e\in E_X$ appears in the path at most twice. 
The bound (\ref{alb}) then follows. 
\qed 
 
We give, now, a recursive definition that will be used to 
parametrize the exponential decay of the potential at different scales. 
Recall definitions (\ref{cjp->Ys}) and (\ref{treedec0}), set  
\begin{equation} 
\label{alb_iY} 
\begin{array}{ll} 
{\displaystyle 
\cT_0(G,s):= 
          \inf_{x\in\dEs_s(G)} 
          \inf_{\newatop{x_m\in g_m}{m=1,\dots,n}} 
          \tree\(\{x,x_1,\dots,x_n\}\)} 
& 
\textrm{ for } 
G=\{g_1,\dots,g_n\}\subset\subset\cG_{\ge 1}\textrm{ and } s\ge 0\\ 
&\\ 
{\displaystyle 
\cT_j(G,s):= 
          \inf_{\ul{R}\in\ul{\cR}_j(G,s)}\; 
          \sum_{R\in\ul{R}}\;\sum_{(H,u)\in R}\, 
          \cT_{j-1}(H,u)} 
& 
\textrm{ for } 
j\ge 1,\, G\subset\subset\cG_{\ge j+1},\textrm{ and } s\ge 0\\ 
\end{array} 
\end{equation} 
As usual if $\underline{R}_j(G,s)=\emptyset$ we understand 
$\cT_j(G,s)=+\infty$. Note that $\cT_0(G,0)=\cT(G)$, see (\ref{alb_iG}). 
Finally for each $j\ge 0$, 
$G\subset\subset\cG_{\ge j+1}$ and $s\ge 0$ we set 
$\hat{\cT}_j(G,s):=\inf_{0\le k\le j}\cT_k(G,s)$. 
In order to clarify the recursive definition (\ref{alb_iY}) we consider  
in some detail the case $j=1$, $G=\{g_1,g_2\}\subset\cG_2$, and $s=0$. 
Let $\underline{R}^*\in\underline{\cR}_1(\{g_1,g_2\},0)$ be a minimizer  
for the right--hand side of (\ref{alb_iY}). Then  
$$ 
\cT_1(\{g_1,g_2\},0)= 
 \sum_{R\in\underline{R}^*} 
 \sum_{(H,u)\in R} 
  \cT_0(H,u) 
$$ 
We note that a polymer $R\in\underline{R}^*$ is built of bonds $(H,u)$ 
connecting on $1$--gentle atoms. Therefore, $\cT_1(\{g_1,g_2\},0)$ 
can be strictly smaller than $\disuno(g_1,g_2)$ due to the presence of  
$1$--gentle atoms between $g_1$ and $g_2$. However, by the sparseness 
conditions \ref{gent:diam} and \ref{gent:dist} of  
Definition~\ref{t:gentle}, we have  
$$ 
\cT_1(\{g_1,g_2\},0)\ge 
 \frac{\gamma_1}{\Gamma_1+\gamma_1}\,\disuno(g_1,g_2)\ge 
 \Big(1-\frac{\Gamma_1}{\gamma_1}\Big)\,\disuno(g_1,g_2) 
$$ 
Indeed, the maximum number of $1$--gentle atoms that can be  
arranged between $g_1$ and $g_2$ is $\disuno(g_1,g_2)/(\Gamma_1+\gamma_1)$. 
The following proposition states a similar bound for a general situation.  
 
\bpro{t:nausea} 
Let $j\ge 0$, $G\ssu\cG_{\ge j+1}$, and $s\ge 0$. 
Then  
\begin{equation} 
\label{nausea} 
\cT_j(G,s)\ge 
          \Big(1-\sum_{k=0}^j\frac{\Gamma_k}{\gamma_k}\Big) 
	  \Big\{\cT(G)+ 
	     \id_{s\ge1}\big[\disuno(\env(\proj{G}),\dEs_s(G))-\vartheta_j\big] 
          \Big\} 
\end{equation} 
where we understand $0/\gamma_0=0$ even if $\gamma_0=0$. 
\end{pro} 
 
We remark that from the bound (\ref{nausea}) above, item~\ref{seq:covol} 
in Definition~\ref{t:seq}, and (\ref{esg0}) 
it is straightforward to deduce that  
\begin{equation} 
\label{alb_iY2} 
\hat\cT_j(G,s)\ge 
          \frac{1}{2}\cT(G)+ 
	  \frac{1}{4}\disuno(\env(\proj{G}),\dEs_s(G)) 
\end{equation} 
 
To prove Proposition~\ref{t:nausea} one of the ingredients is a lemma  
about one--side projections of graphs to hyper--planes. In order to state it 
we need a few more definitions. Let $\hat n\in\{e_i,-e_i,\,i=1,\dots,d\}$  
be a coordinate direction and $c\in\bN$ an integer; we consider  
the hyper--plane  
$\pi\equiv\pi_{\hat{n},c}:=\{x\in\bL,\,(x-c\hat{n})\cdot\hat{n}=0\}\subset\bL$,  
where $\cdot$ denotes the canonical inner product in $\bR^d$.  
We then define the half--lattices  
$\bL_{\pi,\le}:=\{x\in\bL,\,(x-c\hat{n})\cdot\hat{n}\le0\}$ and 
$\bL_{\pi,>}:=\{x\in\bL,\,(x-c\hat{n})\cdot\hat{n}>0\}$; remark  
that $\bL_{\pi,\le}\supset\pi$.  
 
Given a connected graph $(V,E)\subset\subset(\bL,\bE)$, recall the definition  
above (\ref{treedec0}), we define  
$V_{\pi,\le}:=V\cap\bL_{\pi,\le}$,  
$V_{\pi,>}:=V\cap\bL_{\pi,>}$,  
$E_{\pi,\le}:=\{e\in E,\,e\subset\bL_{\pi,\le}\}$, and  
$E_{\pi,>}:=\{e\in E,\,e\cap\bL_{\pi,>}\neq\emptyset\}$. We note that  
$V=V_{\pi,\le}\cup V_{\pi,>}$ and  
$E=E_{\pi,\le}\cup E_{\pi,>}$. We finally define  
$E_{\pi,>}^\perp:=\big\{\{x,y\}\subset\pi, 
                            \,\exists k\ge1\textrm{ such that } 
			    \{x+k\hat{n},y+k\hat{n}\}\in E_{\pi,>}\big\}$. 
                                     
\begin{lem} 
\label{t:prenausea} 
Let $(V,E)\subset(\bL,\bE)$ be a connected graph,  
$\hat n\in\{e_i,-e_i,\,i=1,\dots,d\}$ a coordinate direction, 
and $c\in\bN$; consider the hyper--plane  
$\pi_{\hat{n},c}\equiv\pi\subset\bL$.  
With the definitions given above, if  
$V_{\pi,\le}\neq\emptyset$ then  
\begin{enumerate} 
\item 
\label{i:pren-1} 
the bound  
\begin{equation} 
\label{prenausea} 
|E|\ge|E_{\pi,\le}\cup E_{\pi,>}^\perp|+ 
      \sup_{v\in V_{\pi,>}}\disuno(v,\pi) 
\end{equation} 
holds,  
where we understand the second term in the right--hand side equal to  
zero whenever $V_{\pi,>}=\emptyset$; 
\item 
\label{i:pren-2} 
the pair $(V_{\pi,\le},E_{\pi,\le}\cup E_{\pi,>}^\perp)$ is a  
connected graph. 
\end{enumerate} 
\end{lem} 
 
We remark that this lemma depends on the use of the distance $\disuno$ in the  
definition of the edge set $\bE$. Indeed it would have been false   
if we had used the distance $\disinfinito$.  
 
\smallskip 
\noi{\it Proof of Lemma~\ref{t:prenausea}.\/}\ \ 
Proof of the item~\ref{i:pren-1}.  
Let  
$E_{\pi,>}^\parallel:=\big\{\{x,y\}\in E_{\pi,>}, 
                            \,(y-x)\cdot\hat{n}\neq0\big\}$; 
it is immediate to show that  
\begin{equation} 
\label{pren-1} 
|E|\ge|E_{\pi,\le}\cup E_{\pi,>}^\perp|+|E_{\pi,>}^\parallel| 
\end{equation} 
If $V_{\pi,>}=\emptyset$ (\ref{prenausea}) trivially  
follows from (\ref{pren-1}). Suppose, now,  $V_{\pi,>}\neq\emptyset$. 
Pick $v\in V_{\pi,>}$ and let $D:=\disuno(\pi,v)=\disinfinito(\pi,v)$. 
Recalling that the graph $(V,E)$ is connected and that  
by hypotheses $V_{\pi,\le}\neq\emptyset$, we have that  
there exist $w\in\pi$ and a connected path  
$\ell_1,\dots,\ell_h$ such that  
$v\in\ell_1$, $w\in\ell_h$, and $\ell_m\in E_{\pi,>}$ for  
all $m=1,\dots,h$. We have the obvious bounds 
\begin{equation} 
\label{pren-2} 
\big|E_{\pi,>}^\parallel\big|\ge 
\big|\{\{x,y\}\in\{\ell_1,\dots,\ell_h\},\,(y-x)\cdot\hat{n}\neq0\}\big| 
\ge D 
\end{equation} 
The inequality (\ref{prenausea}) follows from (\ref{pren-1}) and  
(\ref{pren-2}).  
 
Proof of the item~\ref{i:pren-2}. The statement is trivial if 
$|V_{\pi,\le}|=1$.  
Suppose, now, $|V_{\pi,\le}|\ge2$ and pick two distinct vertexes 
$v,w\in V_{\pi,\le}$. By recalling that  
$(V,E)$ is a connected graph we have that there exists 
a connected path joining $v$ to $w$ namely, there exist  
$\ell_1,\dots,\ell_h\in E$ such that  
$v\in\ell_1$, $w\in\ell_h$, and $\ell_m\cap\ell_{m+1}\neq\emptyset$ for  
$m=1,\dots,h-1$.  
 
We let $\ell'_1,\dots,\ell'_{h'}$ be the path obtained from  
$\ell_1,\dots,\ell_h$ by removing all the edges belonging to  
$E_{\pi,>}^\parallel$; we remark that the path $\ell'_1,\dots,\ell'_{h'}$  
is not necessarily connected and that $1\le h'\le h$.  
Let $\ell'=\{x',y'\}$ be an edge of such a path; $\ell'$ 
is either in $E_{\pi,\le}$   
or in $E_{\pi,>}\setminus E_{\pi,>}^\parallel$.  
We set $\bar\ell':=\ell'$ in the former case and  
$\bar\ell':=\{x'+(c-x'\cdot\hat{n})\hat{n}, y'+(c-y'\cdot\hat{n})\hat{n}\} 
 \in E_{\pi,>}^\perp$ 
in the latter. 
 
By construction  
$\bar\ell'_1,\dots,\bar\ell'_{h'}\in E_{\pi,\le}\cup E_{\pi,>}^\perp$.  
Moreover it is an easy task to prove that  
$v\in\bar\ell'_1$, $w\in\bar\ell'_{h'}$, and   
$\ell'_m\cap\ell'_{m+1}\neq\emptyset$ for $m=1,\dots,h'-1$.  
The proof of item~\ref{i:pren-2} is completed. 
\qed 
 
\begin{lem} 
\label{t:passozero} 
Let $G\subset\subset\cG_{\ge1}$ and $s\ge0$. Then the bound (\ref{nausea}) 
holds true for $j=0$.  
\end{lem} 
 
\noi{\it Proof.\/}\  
The statement is trivial in the case $s=0$.  
Let $s\ge1$ and label the elements of $G$ by setting  
$G=\{g_1,\dots,g_{|G|}\}$.  
Let $x^*\in\dEs_s(G)$, $x^*_1\in g_1$, $\dots$, $x^*_{|G|}\in g_{|G|}$  
be a minimizer for the infimum in the definition 
of $\cT_0(G,s)$, see (\ref{alb_iY}). Let also  
$V:=\{x^*, x^*_1,\dots, x^*_{|G|}\}$ and $(V,E)$ the connected graph  
such that  
$|E|=\bT\big(\{x^*, x^*_1,\dots, x^*_{|G|}\}\big)=\cT_0(G,s)$. 
 
Let $F_{x^*}$ the face of $\dEs_s(G)$ such that $x^*\in F_{x^*}$ (choose  
anyone if it is not unique) and $\pi$ the hyper--plane parallel to  
$F_{x^*}$ such that $\pi\cap\env(\proj{G})\neq\emptyset$ and  
$\disuno(\pi,F_{x^*})$ is minimal.  
Let also $\hat{n}$ be the normal to $\pi$ such that $(x^*-y)\cdot\hat{n}>0$  
for any $y\in\pi$.   
By applying the Lemma~\ref{t:prenausea} to the graph  
$(V,E)$, the normal $\hat{n}$, and the hyper--plane $\pi$ we get  
$ 
|E|\ge|E_{\pi,\le}\cup E_{\pi,>}^\perp|+\disuno(x^*,\pi) 
$. 
Since $V_{\pi,\le}=\{x^*_1,\dots,x^*_{|G|}\}$, by item~\ref{i:pren-2} 
of the Lemma~\ref{t:prenausea} we have that  
$ 
|E_{\pi,\le}\cup E_{\pi,>}^\perp| 
 \ge\bT\big(\{x^*_1,\dots,x^*_{|G|}\}\big) 
 \ge\cT(G) 
$. 
Moreover, by construction $\disuno(x^*,\pi)=\disuno(\env(\proj{G}),\dEs_s(G))$. 
The thesis follows. 
\qed 
 
\smallskip 
\noi{\it Proof of Proposition~\ref{t:nausea}.\/}\ \ 
We can assume $\underline{R}_j(G,s)\neq\emptyset$, otherwise  
$\cT_j(G,s)=+\infty$. 
We prove (\ref{nausea}) by induction;  
the step $j=0$ has been proven in the Lemma~\ref{t:passozero}. 
We suppose (\ref{nausea}) holds for $j-1$ and we show it holds true  
for $j$. 
To bound $\cT_j(G,s)$ we let 
$\ul{R}^*\in\ul{\cR}_j(G,s)$ be a minimizer for (\ref{alb_iY}).  
Note that $\ul{R}_j(G,s)$ is not a finite set because repetitions of  
the some bond are allowed. However a minimizer $\underline{R}^*$  
does exist because without such repetitions $\ul{R}_j(G,s)$ would be  
finite and {\em repetita juvant}. 
We have 
\begin{equation} 
\label{potryo} 
\cT_j(G,s)=\sum_{R\in\ul{R}^*}\sum_{(H,u)\in R}\cT_{j-1}(H,u) 
\end{equation} 
 
We consider, now,  the case $s=0$.  
Let  
$\cH\equiv\cH(\ul{R}^*):= 
  \{H\subset\cG_{\ge j}:\,\exists R\in\ul{R}^*,\,\exists u\ge 0:\, 
    (H,u)\in R\textrm{ and } |H|\ge 2\}$; 
we note that $\cH$ is finite and not empty. 
From (\ref{potryo}) and the inductive hypothesis we have 
$$ 
\cT_j(G,0)\ge 
          \Big(1-\sum_{k=0}^{j-1}\frac{\Gamma_k}{\gamma_k}\Big) 
          \sum_{H\in\cH} 
          \cT(H) 
$$ 
We also remark that definitions (\ref{cRi}) and (\ref{cjp->Ys}) 
imply that the system $\cH$ is $j$--connected in the sense specified just 
above (\ref{polii}). 
By adding and subtracting 
$\Gamma_j/\gamma_j$ and by remarking that $|H|\ge 2$ implies 
$\cT(H)\ge\gamma_j$ we get 
\begin{equation} 
\label{cac3} 
\cT_j(G,0) 
          \ge 
          \Big(1-\sum_{k=0}^{j}\frac{\Gamma_k}{\gamma_k}\Big) 
          \sum_{H\in\cH} 
          \cT(H) 
          +|\cH|\Gamma_j 
\end{equation} 
Let us construct a partition of the system $\cH$: pick  
an element of $\cH$, denote it by 
$H_{0,1}$, and set $\cH_0:=\{H_{0,1}\}$. For any $m\ge 1$ and 
$H\in\cH\setminus\bigcup_{\ell=0}^{m-1}\cH_{\ell}$ we say that 
$H\in\cH_m$ if and only if there exists $H'\in\cH_{m-1}$ such that 
$H$ and $H'$ are $j$--connected namely, 
$H\cap H'\cap\cG_j\neq\emptyset$. Recalling 
$\cH$ is $j$--connected we have that 
there exists a maximal value of $m$ that we call $t$; in other words 
there exists $t\ge 0$ such that 
$\cH_m\neq\emptyset$ for all $m\le t$ and $\cH_m=\emptyset$ 
for all $m>t$. The collection $\cH_0,\dots,\cH_t$ is a partition 
of $\cH$. 
 
{}For each $m=1,\dots,t$ we denote by $H_{m,1},\dots,H_{m,|\cH_m|}$ the 
elements of $\cH_m$; for each $m=0,\dots,t$ and $\ell=1,\dots,|\cH_m|$ 
we let $(V_{m,\ell},E_{m,\ell})\subset(\bL,\bE)$ be a connected graph such that 
\begin{equation} 
\label{cac5.1} 
\cT(H_{m,\ell})=|E_{m,\ell}| 
\end{equation} 
and for each $h\in H_{m,\ell}$ we have that $V_{m,\ell}\cap h\neq\emptyset$. 
We define, now, an algorithm that constructs 
a graph $(V,E)\subset(\bL,\bE)$  
such that $|E|\ge\cT(G)$ and $|E|$ is bounded from above in terms of 
$\cT(H)$ for $H\in\cH$: 
\texttt{ 
\begin{enumerate} 
\item 
\label{i:al1} 
set $m=0$ and $(V,E)=(V_{0,1},E_{0,1})$; 
\item 
\label{i:al2} 
set $m=m+1$ and $\ell=0$, if $m=t+1$ goto \ref{i:al6}; 
\item 
\label{i:al3} 
set $\ell=\ell+1$, pick $\ell'\in\{1,\dots,|\cH_{m-1}|\}$ such that 
$H_{m-1,\ell'}\connj H_{m,\ell}$; 
\item 
\label{i:al3.1} 
pick 
$h\in H_{m-1,\ell'}\cap H_{m,\ell}\cap\cG_j$, 
$y\in h\cap V_{m-1,\ell'}$, and 
$x\in h\cap V_{m,\ell}$; 
\item 
\label{i:al4} 
find a connected graph $(W,F)\subset(\bL,\bE)$ such that $|F|$ is minimal  
and the set of vertices $W$ contains both $x$ and $y$; 
\item 
\label{i:al4.1} 
set $V=V\cup V_{m,\ell}\cup W$ and $E=E\cup E_{m,\ell}\cup F$; 
\item 
\label{i:al5} 
if $\ell< |\cH_m|$ goto \ref{i:al3} else goto \ref{i:al2}; 
\item 
\label{i:al6} 
exit; 
\end{enumerate} 
}\noindent 
By recursion it is easy to prove that this algorithm 
outputs a connected graph $(V,E)$ such that for each $H\in\cH$ and $h\in H$ 
there exists $x\in  h$ such that $x\in V$; 
in particular for each $g\in G$ 
there exists $x\in  g$ such that $x\in V$, hence $|E|\ge\cT(G)$. 
Moreover, by noticing 
that the graph $(W,F)$ introduced at line \ref{i:al4} is such that 
$|F|\le\diamuno(h)\le\Gamma_j$ we have 
\begin{equation} 
\label{cac4} 
|E|\le\sum_{m=0}^t\sum_{\ell=1}^{|\cH_m|}|E_{m,\ell}|+(|\cH|-1)\Gamma_j 
\end{equation} 
Now, by using (\ref{cac3})--(\ref{cac4}) we get 
$$ 
\cT_j(G,0) 
          \ge 
          \Big(1-\sum_{k=0}^{j}\frac{\Gamma_k}{\gamma_k}\Big) 
          \big[|E|-(|\cH|-1)\Gamma_j\big] 
          +|\cH|\Gamma_j 
	  \ge 
          \Big(1-\sum_{k=0}^{j}\frac{\Gamma_k}{\gamma_k}\Big) 
	  \cT(G) 
$$ 
which completes the inductive proof of (\ref{nausea}) for $s=0$. 
 
We consider, now, the case $s\ge1$.  
Recalling (\ref{potryo}),  
there exists $R'\in\ul{R}^*$ and $(H',u')\in R'$ such that 
$\Es_{u'}(H')\cap\dEs_s(G)\neq\emptyset$.  
Let 
$\cH'\equiv\cH'(\ul{R}^*):= 
  \{H\subset\cG_{\ge j}:\,\exists R\in\ul{R}^*,\,\exists u\ge 0:\, 
    (H,u)\in R,\, (H,u)\neq(H',u')\textrm{ and } |H|\ge 2\}$. 
Note that, as in the previous case, $|H|\ge 2$ implies 
$\cT(H)\ge\gamma_j$; on the other hand we note that $\cH'$ can be empty. 
Set also $\cH:=\cH'\cup\{H'\}$.  
By using (\ref{potryo}) and the recursive hypothesis we have 
\begin{equation} 
\label{cac6} 
\begin{array}{rl} 
\cT_j(G,s)\ge & 
         {\displaystyle 
          \cT_{j-1}(H',u')+\sum_{H\in\cH'}\cT_{j-1}(H,u) 
         }\\ 
          \ge & 
         {\displaystyle 
          \Big(1-\sum_{k=0}^{j-1}\frac{\Gamma_k}{\gamma_k}\Big) 
          \Big\{\id_{u'\ge1}\big[\disuno(\env(\proj{H'}),\dEs_{u'}(H')) 
                                 -\vartheta_{j-1}\big] 
               +\sum_{H\in\cH}\cT(H) 
          \Big\} 
         }\\ 
\end{array} 
\end{equation} 
We note that for each $H\in\cH'$ we have $|H|\ge2$, hence  
$\cT(H)\ge\gamma_j$. Moreover, we claim that  
\begin{equation} 
\label{cac16} 
\cT(H')+\id_{u'\ge1}\big(\disuno(\env(\proj{H'}),\dEs_{u'}(H')) 
                         -\vartheta_{j-1}\big) 
\ge\gamma_j 
\end{equation} 
Indeed, if $u'=0$ then $|H'|\ge2$, so that $\cT(H')\ge\gamma_j$. 
On the other end if $u'\ge1$, then 
$\disuno(\env(\proj{H'}),\dEs_{u'}(H'))=\vartheta_j+u'$ implies  
$ 
\disuno(\env(\proj{H'}),\dEs_{u'}(H'))-\vartheta_{j-1}> 
 \vartheta_j-\vartheta_{j-1}=\Gamma_j+\gamma_j>\gamma_j 
$. 
Now, by adding and subtracting $\Gamma_j/\gamma_j$ in   
(\ref{cac6}) we get  
\begin{equation} 
\label{cac10} 
\cT_j(G,s)\ge 
          \Big(1-\sum_{k=0}^j\frac{\Gamma_k}{\gamma_k}\Big) 
          \Big\{\sum_{H\in\cH}\cT(H) 
	       +\id_{u'\ge1}\big[\disuno(\env(\proj{H'}),\dEs_{u'}(H')) 
                                 -\vartheta_{j-1}\big] 
	  \Big\} 
          +|\cH|\Gamma_j 
\end{equation} 
 
Since $\Es_{u'}(H')\cap\dEs_s(G)\neq\emptyset$ there exists 
$h'\in H'$ such that $\disuno(h',\dEs_s(G))=\vartheta_j+u'$. 
Label the elements of $G$ by setting $G=\{g_1,\dots,g_{|G|}\}$. 
By running the algorithm used in the case $s=0$,  
we construct a connected graph $(V,E)\subset(\bL,\bE)$ such that 
$V\supset\{x',x_1,\dots,x_{|G|}\}$, for some  
$x'\in h'$, $x_1\in g_1,\dots,x_{|G|}\in g_{|G|}$, and 
\begin{equation} 
\label{cac8} 
\sum_{H\in\cH}\cT(H)\ge |E|-(|\cH|-1)\Gamma_j 
\end{equation} 
Let $F'$ be the face of $\dEs_s(G)$ such that $\disuno(h',F')=\vartheta_j+u'$ 
(choose anyone if it is not unique) and  
$\pi$ the hyper--plane parallel to  
$F'$ such that $\pi\cap\env(\proj{G})\neq\emptyset$ and  
$\disuno(\pi,F')$ is minimal.  
Let also $\hat{n}$ be the normal to $\pi$ such that $(y'-y)\cdot\hat{n}>0$  
for any $y'\in F'$ and $y\in\pi$.   
By applying the Lemma~\ref{t:prenausea} to the graph  
$(V,E)$, the normal $\hat{n}$, and the hyper--plane $\pi$ we get  
\begin{equation} 
\label{cac9} 
|E| 
\ge 
 \cT(G)+\disuno(\env(\proj{G}),h') 
\end{equation} 
Finally, by plugging (\ref{cac8}) and (\ref{cac9}) into (\ref{cac10}) we get  
\begin{equation} 
\label{cac18} 
\cT_j(G,s)\ge 
          \Big(1-\sum_{k=0}^j\frac{\Gamma_k}{\gamma_k}\Big) 
          \Big\{\cT(G)+\disuno(\env(\proj{G}),h')+\Gamma_j 
	       +\id_{u'\ge1}\big[\disuno(\env(\proj{H'}),\dEs_{u'}(H')) 
                                 -\vartheta_{j-1}\big] 
	  \Big\} 
\end{equation} 
 
Consider, now, the sub--case $u'=0$.  
In this case $\disuno(h',F')=\vartheta_j$, hence $h'\not\subset\env(\proj{G})$. 
This implies   
$h'\in\cG_j$; therefore $\diamuno(h')\le\Gamma_j$, see  
item~\ref{gent:diam} in Definition~\ref{t:gentle}.  
We get  
\begin{equation} 
\label{cac11} 
\disuno(\env(\proj{G}),h') 
\ge 
 \disuno(\env(\proj{G}),\dEs_s(G))-\Gamma_j-\vartheta_j 
\end{equation} 
The bound (\ref{nausea}) follows from (\ref{cac18}) and (\ref{cac11}). 
 
We finally consider the sub--case $u'\ge1$.  
Recalling how $h'\in H'$ has been chosen, we have that  
\begin{equation} 
\label{cac20} 
\disuno(\env(\proj{H'}),\dEs_{u'}(H')) 
 =\disuno(h',\dEs_{u'}(H')) 
 =\disuno(h',\dEs_{s}(G)) 
\end{equation} 
If $h'\in G$ then $h'\subset\env(\proj{G})$; hence  
$\disuno(\env(\proj{H'}),\dEs_{u'}(H'))\ge\disuno(\env(\proj{G}),\dEs_{s}(G))$. 
Then (\ref{nausea}) follows easily from (\ref{cac18}).  
On the other hand if $h'\in\cG_j$, we have $\diamuno(h')\le\Gamma_j$, hence 
by using (\ref{cac20}) we have 
\begin{equation} 
\label{cac22} 
\disuno(\env(\proj{G}),h')+\Gamma_j+\disuno(\env(\proj{H'}),\dEs_{u'}(H')) 
\ge 
\disuno(\env(\proj{G}),\dEs_s(G)) 
\end{equation} 
Then (\ref{nausea}) follows easily from (\ref{cac18}).  
\qed  
 
\begin{lem} 
\label{t:dgg'} 
Let $j\ge 0$, $G\subset\subset\cG_{\ge j+1}$, $s\ge 0$; 
suppose $\ul{\cR}_j(G,s)\neq\emptyset$,  
see definition (\ref{cjp->Ys}). For each $g\in G$, 
$\ul{R}\in\ul{\cR}_j(G,s)$ and $h\in\ul{R}\rest_j$; we have 
\begin{equation} 
\label{dgg'} 
\sum_{R\in\ul{R}}\sum_{(H,u)\in R} 
 \hat\cT_{j-1}(H,u) 
\ge 
 \frac{1}{2}\dis(g,h) 
\end{equation} 
\end{lem} 
 
\Pro 
The Lemma can be proven by using   
(\ref{nausea}), the simple bound 
$1-\sum_0^{\infty}(\Gamma_j/\gamma_j)\ge 1/2$, and 
by running the algorithm introduced in the proof of Lemma~\ref{t:nausea}. 
\qed 
 
\blem{t:distre} 
Let $G\ssu\cG_{\ge j+1}$, $s\ge 0$ and $\ul{\cR}_j(G,s)$ as defined in 
\eqref{cjp->Ys}. Then, for each $\ul{R}\in \ul{\cR}_j(G,s)$, 
\be{distre} 
\sum_{R\in\ul{R}} \: \sum_{(H,u)\in R} 
|H| \ge  |G| + \sum_{R\in\ul{R}} \lmo R\rest_{j} \rmo 
\end{equation} 
\elem 
 
\Pro 
The Lemma follows directly from the definition of 
$\ul{\cR}_j(G,s)$. 
\qed 
 
\subsec{Preliminary lemmata}{s:prlemce} 
\par\noindent 
In this section we collect some technical bounds needed to prove the 
convergence of the multi--scale cluster expansion. 
\blem{t:elbound} 
For $m>0$ let 
\be{dKm} 
K(m):= \Big( 
       \frac{1+e^{-m/2}}{1-e^{-m/2}} 
       \Big)^d 
\end{equation} 
where we recall $d$ is the dimension of the lattice $\bL$.  
Let also $\gamma,L\ge0$ be positive reals; then 
we have 
\be{elbound} 
\sum_{\newatop{x\in\bL:} 
               {\disuno(x,0)\ge\gamma}} 
e^{-m\,\dis\(x,B_{L}\)} 
\le K(m)\,e^{-\frac{m}{2}(\gamma-2L)} 
\end{equation} 
where we recall $B_L$ is the ball of radius $L$ centered at the origin  
defined at the end of Section~\ref{s:lat}. 
\elem 
 
\Pro 
First of all we note that  
$\disuno(x,0)\le L+\disuno(x,B_L)$. Hence 
$$ 
\sum_{\newatop{x\in\bL:}{\disuno(x,0)\ge\gamma}}e^{-m\,\dis(x,B_{L})} 
\le 
\sum_{\newatop{x\in\bL:}{\disuno(x,0)\ge\gamma}}e^{-m\,[\dis(x,0)-L]} 
\le 
e^{mL-m\gamma/2} 
\sum_{\newatop{x\in\bL:}{\disuno(x,0)\ge\gamma}}e^{-m\dis(x,0)/2} 
$$ 
Recalling that $\disuno(x,0)=|x_1|+\cdots+|x_d|$, where $x=(x_1,\dots,x_d)$, 
and using the bound above we get  
$$ 
\sum_{\newatop{x\in\bL:}{\disuno(x,0)\ge\gamma}}e^{-m\,\dis(x,B_{L})} 
\le 
e^{-m(\gamma-2L)/2} 
\sum_{x\in\bL}e^{-m(|x_1|+\cdots+|x_d|)/2} 
\le 
e^{-m(\gamma-2L)/2}\Big(1+2\sum_{k=1}^\infty e^{-m\,k/2}\Big)^d 
$$ 
and the Lemma follows via elementary computations. 
\qed 
 
\blem{t:stepj} 
For $j\ge 1$ and $m>0$ let 
\be{qj} 
q_j(m):=K(m/4)\,e^{-m\,\gamma_j/8} 
\end{equation} 
where $K(m)$ has been defined in Lemma~\ref{t:elbound}. 
Assume $q_j(m)<1$ and set 
\be{Kj} 
K_j(m):= 
  e^{-\frac{m}{4}\vartheta_j} 
      \frac{e^{-m/4}}{1-e^{-m/4}}+ 
   \Big[1+e^{-\frac{m}{4}\vartheta_j} 
      \frac{e^{-m/4}}{1-e^{-m/4}}\Big] 
  (\Gamma_j+1)^d\frac{q_j(m)}{1-q_j(m)} 
\end{equation} 
Then 
\be{stepj} 
\sup_{g\in\cG_j} \sum_{\newatop{G\ssu\cG_{\ge j}:} 
                               {G\ni g}} 
\sum_{s=0}^\infty \id_{(|G|,s)\neq(1,0)} 
\exp\big\{-m\hat{\cT}_{j-1} (G,s)\big\} 
\le K_j(m) 
\end{equation} 
\elem 
 
\Pro 
Let $g_0\in\cG_j$; by using (\ref{alb_iY2}), definition 
(\ref{alb_iG}), and Lemma~\ref{t:infpi} we have 
\bea{j1} 
&& 
\sum_{\newatop{G\ssu\cG_{\ge j}} 
             {G\ni g_0}} 
\sum_{s=0}^\infty \id_{(|G|,s)\neq(1,0)} 
\exp\big\{-m\hat{\cT}_{j-1}(G,s)\big\} 
\nn\\ 
&&\phantom{mer} 
\le 
\sum_{s=1}^\infty 
\exp\Big\{-\frac{m}{4}\disuno\(\env(g_0),\dEs_s(g_0)\)\Big\} 
\nn\\ 
&&\phantom{merda} 
+ \: 
\sum_{k=1}^\infty 
\sum_{\newatop{G\ssu\cG_{\ge j}:\,G\ni g_0} 
              {|G|=k+1}} 
\exp\Big\{-\frac{m}{4} 
\inf_{\newatop{x_h\in g_h:} 
              {h=0,1,\dots,k}} 
\inf_{\pi\in\Pi_0(k)} 
\sum_{l=1}^{k}\disuno\left( x_{\pi(l-1)},x_{\pi(l)}\right) 
\Big\} 
\nn\\ 
&&\phantom{merdone} 
\times \: \sum_{s=0}^\infty 
\exp\Big\{-\frac{m}{4}\disuno\(\env(\proj{G}),\dEs_s(G)\)\Big\} 
\end{eqnarray} 
For $G\subset\subset\cG_{\ge j}$, such that $G\cap\cG_j\neq\emptyset$, 
we have $\disuno(\env(\proj{G}),\dEs_s(G))=\vartheta_j+s$; then 
\begin{equation} 
\label{j3.1} 
\sum_{s=1}^\infty 
 e^{-\frac{m}{4}\disuno\(\env(g_0),\dEs_s(g_0)\)} 
 =e^{-\frac{m}{4}\vartheta_j} 
  \frac{e^{-m/4}}{1-e^{-m/4}} 
\end{equation} 
and 
\begin{equation} 
\label{j3} 
\sum_{s=0}^\infty 
 e^{-\frac{m}{4}\disuno\(\env(\proj{G}),\dEs_s(G)\)} 
 =1+e^{-\frac{m}{4}\vartheta_j} 
    \frac{e^{-m/4}}{1-e^{-m/4}} 
\end{equation} 
On the other hand 
\bea{j4} 
&& 
\sum_{k=1}^\infty 
\sum_{\newatop{G\su\cG_{\ge j}:\,G\ni g_0} 
              {|G|=k+1}} 
\exp\Big\{-\frac{m}{4} 
\inf_{\newatop{x_h\in g_h} 
              {h=0,1,\dots, k}} 
\inf_{\pi\in\Pi_0(k)} 
\sum_{h=1}^{k} \disuno(x_{\pi(h-1)},x_{\pi(h)}) 
\Big\} 
\nn\\ 
&&\phantom{mer} 
\le 
\sum_{k=1}^\infty \frac {1}{k!} 
\sum_{\newatop{g_1,\dots,g_k\in\cG_{\ge j}:} 
              {g_h\neq g_{h'},\, g_h\neq g_0}} 
\exp\Big\{-\frac{m}{4} 
\inf_{\newatop{x_h\in g_h:} 
              {h=0,1,\dots,k}} 
\inf_{\pi\in\Pi_0(k)} 
\sum_{h=1}^{k}\disuno(x_{\pi(h-1)},x_{\pi(h)}) 
\Big\} 
\nn\\ 
&&\phantom{mer} 
\le 
\sum_{k=1}^\infty \frac {1}{k!} 
\sum_{\newatop{g_1,\dots,g_k\in\cG_{\ge j}:} 
              {g_h\neq g_{h'},\,g_h\neq g_0}} 
\sum_{\newatop{x_h\in g_h:} 
              {h=0,1,\dots, k}} 
\sum_{\pi\in\Pi_0(k)} 
\exp\Big\{-\frac{m}{4} 
\sum_{h=1}^{k}\disuno(x_{\pi(h-1)},x_{\pi(h)}) 
\Big\} 
\nn\\ 
&&\phantom{mer} 
\le 
\sum_{k=1}^\infty \frac {1}{k!} 
\sum_{x_0\in g_0} 
\sum_{\pi\in\Pi_0(k)} 
\prod_{h=1}^k 
\bigg( 
\sum_{\newatop{g_{\pi(h)}} 
              {g_{\pi(h)}\neq g_{\pi(h-1)}}} 
\!\!\!\!\sum_{x_{\pi(h)}\in g_{\pi(h)} } 
\exp\Big\{-\frac{m}{4}\disuno(x_{\pi(h-1)},x_{\pi(h)}) 
\Big\}\bigg) 
\nn\\ 
\end{eqnarray} 
We now have 
\begin{equation} 
\label{j5} 
\begin{array}{rl} 
{\displaystyle 
\sup_{g\in \cG_{\ge j}}\:\sup_{x\in g}\, 
\sum_{\newatop{g'\in \cG_{\ge j}} 
              {g'\neq g}} 
\sum_{y\in g'} 
\exp\Big\{-\frac{m}{4}\disuno(x,y)\Big\}} 
& 
{\displaystyle 
\le\sup_{x\in\bL} 
\sum_{\newatop{y\in\bL} 
              {\disuno(x,y)>\g_j}} 
\exp\Big\{-\frac{m}{4}\disuno(x,y)\Big\}} 
\\ 
&\\ 
& 
{\displaystyle 
\le K(m/4)\exp\Big\{-\frac{m}{8}\g_j\Big\}=q_j(m)}\\ 
\end{array} 
\end{equation} 
where we used Lemma~\ref{t:elbound} and (\ref{qj}). 
 
By plugging \eqref{j5} into the r.h.s.\ of \eqref{j4} we then get 
\bea{j6} 
&& 
\sum_{k=1}^\infty 
\sum_{\newatop{G\su\cG_{\ge j}:\,G\ni g_0} 
              {|G|=k+1}} 
\exp\Big\{-\frac{m}{4} 
\inf_{\pi\in\Pi_0(k)} 
\inf_{\newatop{x_h\in g_h:} 
              {h=0,1,\dots,k}} 
\sum_{h=1}^{k}\disuno(x_{\pi(h-1)},x_{\pi(h)}) 
\Big\} 
\nn\\ 
&&\phantom{mer} 
\le 
\sum_{k=1}^\infty \frac {1}{k!} 
\sum_{x_0\in g_0}  
\sum_{\pi\in\Pi_0(k)}  
[q_j(m)]^k 
\nn\\ 
&&\phantom{mer} 
=|g_0|\,\frac{q_j(m)}{1-q_j(m)} 
\le (\G_j+1)^d\,\frac {q_j(m)}{1-q_j(m)} 
\end{eqnarray} 
The estimate \eqref{stepj} now follows collecting the bounds 
\eqref{j1}--\eqref{j3} and \eqref{j6}. 
\qed 
\smallskip 
 
In the sequel we shall need some elementary 
inequalities relating the sequences $\G$, $\g$ 
to the parameters $\alpha$ and $A$ introduced in Condition~\ref{t:ipotesi}. 
We show how those inequalities are implied by the hypotheses 
of Theorem~\ref{t:mainceds}. 
 
\blem{t:condsupp} 
Suppose the hypotheses of Theorem~\ref{t:mainceds} are satisfied. 
We define the decreasing sequence of positive numbers 
\begin{equation} 
\label{delta} 
\delta_k:=\frac{8d}{a^{1/3}}\,\gamma_k^{-1/3} 
\end{equation} 
for $k\ge 1$. Moreover we set 
\begin{equation} 
\label{cc71} 
m_0=\frac{\alpha}{4},\;\;\;\;\textrm{ and }\;\;\;\; 
m_j:=\frac{\alpha}{4}-4\sum_{k=1}^j\delta_k \;\;\textrm{ for all }j\ge 1 
\end{equation} 
Then 
\begin{enumerate} 
\item\label{cc1} 
for each $j\ge1$ we have $\delta_j\gamma_j\ge 8j$; 
\item\label{cc17} 
we have 
${\displaystyle \sum_{k=1}^{\infty}\delta_k\le\frac{\alpha}{32}}$; 
\item\label{conv:0} 
we have $e\varepsilon<1/3$; 
\item\label{conv:4} 
let $q_j(m)$ as defined in Lemma~\ref{t:stepj} and $\delta_j$ as in 
(\ref{cc17}), then 
$q_j(\delta_j)<1$ for all $j\ge 1$; 
\item\label{conv:3} 
let $K_j(m)$ as defined in Lemma~\ref{t:stepj}, then 
$K_j(\delta_j)<1/3$ for all $j\ge 1$; 
\item\label{conv:5} 
for each $j\ge 1$ we have 
$m_{j-1}\gamma_j\ge m_j\gamma_j\ge 32j$; 
\item\label{conv:7} 
we have 
$K(\delta_j/2) \exp\lg -\frac {\delta_j}{4}\g_j\rg\le 1$ 
for all $j\ge 1$; 
\item\label{conv:6} 
we have 
$K((m_{j-1}-2\delta_j)/2) \exp\lg-\frac{m_{j-1}-2\delta_j}{4}\g_j\rg\le1$ 
for all $j\ge 1$; 
\item\label{conv:8} 
we have 
$[4^d(\Gamma_i+\gamma_j)^d+1]\exp\lg -\frac{\delta_j}{4}\g_i\rg\le 1$ 
for any $1\le j<i$. 
\end{enumerate} 
\elem 
 
\Pro 
Item~\ref{cc1} is an immediate consequence of definition (\ref{delta}) 
and item~\ref{i:dc6} in the hypotheses 
of Theorem~\ref{t:rispot}. 
By definition (\ref{delta}) item~\ref{cc17} is equivalent to 
item~\ref{i:dc5} in the hypotheses of Theorem~\ref{t:rispot}. 
 
Item~\ref{conv:0} is an immediate consequence of the definition of  
$\varepsilon$ in item~\ref{i:main3} of Theorem~\ref{t:mainceds}, 
item~\ref{i:dc3} in the hypotheses 
of Theorem~\ref{t:rispot} and the property $\gamma_1>2\Gamma_1$ (see 
item~\ref{seq:<} in Definition~\ref{t:seq}). 
 
With simple elementary computations,  
one can prove that definition 
(\ref{delta}) implies that the inequality 
\begin{equation} 
\label{intin} 
\frac{22}{3}\, \bigg(\frac{18\gamma_j}{\delta_j}\bigg)^d 
\,e^{-\delta_j\gamma_j/8}\le\frac{1}{6} 
\end{equation} 
holds for all $j\ge 1$; such inequality will 
be useful in the proof of the remaining items. 
Indeed, by using (\ref{delta}) we get that (\ref{intin}) is equivalent to  
$(\gamma_j^2/a)^{2/3}\exp\{-(\gamma_j^2/a)^{1/3}\}\le1$, 
which holds trivially.  
 
Item~\ref{conv:4} is obvious once one has proven 
\begin{equation} 
\label{qjint} 
q_j(\delta_j)\le\frac{1}{7(\Gamma_j+1)^d+1} 
\end{equation} 
for all $j\ge 1$. 
To prove (\ref{qjint}) we first use (\ref{dKm}), (\ref{qj}) 
and recall $\Gamma_j\ge 2$ for all $j\ge 1$, see Definition~\ref{t:seq}; 
we then have  
\begin{equation} 
\label{qjint2} 
[7(\Gamma_j+1)^d+1]q_j(\delta_j) 
\le 
\frac{22}{3}\bigg(\frac{3}{2}\bigg)^d\Gamma_j^dq_j(\delta_j) 
\le 
\frac{22}{3}\bigg(\frac{3}{2}\bigg)^d\Gamma_j^d\, 
\Big(\frac{1+e^{-\delta_j/8}}{1-e^{-\delta_j/8}}\Big)^d 
\,e^{-\delta_j\gamma_j/8} 
\end{equation} 
We note, now, that  
item~\ref{i:dc3} in the hypotheses of Theorem~\ref{t:rispot} 
and definition (\ref{delta}) implies  
$\delta_j\le1$ for all $j\ge 1$. Hence, the term  
$(1+e^{-\delta_j/8})/(1-e^{-\delta_j/8})$ can be bounded from above by  
$24/\delta_j$.  
The inequality (\ref{qjint}) finally follows from (\ref{intin}) once  
we recall $\gamma_j\ge2\Gamma_j$ for all $j\ge1$.

Item~\ref{conv:3}: first note that for $j\ge 1$ 
\begin{equation} 
\label{aa2} 
e^{-\delta_j\vartheta_j/4} 
 \frac{e^{-\delta_j/4}}{1-e^{-\delta_j/4}} 
\le 
e^{-\delta_j\gamma_j/8} 
 \frac{1+e^{-\delta_j/4}}{1-e^{-\delta_j/4}} 
\le 
e^{-\delta_j\gamma_j/8}\frac{12}{\delta_j} 
\le 
\frac{1}{6} 
\end{equation} 
where we used $\vartheta_j\ge\gamma_j$ for all $j\ge 1$,  
inequality (\ref{intin}), and 
$\delta_j\le1$ for all $j\ge 1$. 
By inserting the bounds (\ref{qjint}) and (\ref{aa2}) inside the expression 
of $K_j(\delta_j)$, see definition (\ref{Kj}), we get the desired inequality. 
 
Item \ref{conv:5}: from (\ref{cc71}) and item~\ref{cc17} above we have 
that 
\begin{equation} 
\label{aa3} 
m_j=\frac{\alpha}{4}-4\sum_{k=1}^j\delta_k\ge 
\frac{\alpha}{4}-4\frac{\alpha}{32} 
=\frac{\alpha}{8}\ge 4\delta_j 
\end{equation} 
Hence, $m_{j-1}\gamma_j\ge m_j\gamma_j\ge 4\delta_j\gamma_j\ge 32j>4(j-1)$, 
where we have used item~\ref{cc1} above. 
 
Item~\ref{conv:7} is a straightforward consequence of the 
definition (\ref{dKm}) of $K$ and the inequality (\ref{intin}). 
 
Item~\ref{conv:6}: by using (\ref{aa3}) we have that 
$m_{j-1}-2\delta_j\ge\delta_j$. 
So the thesis follows from item \ref{conv:7} once we note that 
$K(m)$ is a decreasing function of $m\ge0$. 
 
Item~\ref{conv:8} follows easily from (\ref{intin}), using that 
$\Gamma_i\ge7\gamma_j$, see item~\ref{seq:con-dist} in 
Definition~\ref{t:seq}, and $\delta_i\le\delta_j$ for $i>j\ge1$. 
\qed 
 
\subsec{Recursive estimate}{s:convce} 
\par\noindent 
In this section we obtain a 
recursive estimate on the effective interaction due to the integration 
on scale $j$, which is the key step in  the proof 
of Theorem~\ref{t:mainceds}. 
More precisely, recalling $\varepsilon$ and $m_j$ have been defined 
in item~\ref{i:main3} of Theorem~\ref{t:mainceds} 
and in (\ref{cc71}), we shall prove the following bounds. 
 
\bteo{t:bounds} 
Let the hypotheses of Theorem~\ref{t:mainceds} be satisfied. 
For $i\ge 1$ set 
$A_i:=(1\vee A)(8^d+1)\Gamma_i^d$. 
Let also $\Psi^{(i,j)}_{g,\Lambda}$ 
(resp.\ $\Phi^{(i,j)}_{G,s,\Lambda}$) as defined 
in (\ref{autopot}) and (\ref{selfintij}) 
(resp.\ in (\ref{eq:veri-pot}) and (\ref{effiintij})). 
Then for each $i>j\ge 0$, we have 
\bea{recauto} 
\|\Psi^{(i,j)}_{g,\Lambda}\|_{\infty} 
  &\le & A_i 
\quad\quad\forall g\in\cG_i \\ 
\label{recpot} 
\|\F^{(i,j)}_{G,s,\Lambda}\|_{\infty} 
&\le & \e^{|G|} e^{-m_j \hat{\cT}_j (G,s) } 
\quad\quad\forall G\ssu\cG_{\ge i}:\, G\cap\cG_i\ne\es,\,\forall s\ge0 
\phantom{mer} 
\end{eqnarray} 
for any $\Lambda\subset\subset\bL$. 
\eteo 
The Theorem follows by complete induction from 
Lemma~\ref{t:step0} and Proposition~\ref{t:re} below. 
First of all we show 
that \eqref{recauto} and \eqref{recpot} hold for $j=0$. 
 
\blem{t:step0} 
Let $\Ps^{(i,0)}_{g,\Lambda}$, resp. $\F^{(i,0)}_{G,s,\Lambda}$, as defined in 
\eqref{autopot}, resp. in \eqref{eq:veri-pot} 
and assume the hypotheses of Theorem~\ref{t:mainceds} are satisfied. 
Then for any $\Lambda\subset\subset\bL$ and any $i\ge 1$ 
\bea{sauto0} 
\|\Psi^{(i,0)}_{g,\Lambda}\|_{\infty} 
  &\le & A_i 
\quad\quad\forall g\in\cG_i \\ 
\label{sint0} 
\|\F^{(i,0)}_{G,s,\Lambda}\|_{\infty} 
&\le & \e^{|G|} e^{-m_0 \hat{\cT}_0 (G,s) } 
\quad\quad\forall G\ssu\cG_{\ge i}:\, G\cap\cG_i\ne\es,\,\forall s\ge0 
\phantom{mer} 
\end{eqnarray} 
\elem 
 
\Pro 
We first prove \eqref{sauto0}.  
Recall (\ref{autopot}), given $X\in\Upsilon_\Lambda(g,0)$, 
if $\xi(X)=g$ then $X\cap g\neq0$. Hence 
by using Condition~\ref{t:ipotesi}, 
\be{auto01} 
\|\Psi^{(i,0)}_{g,\Lambda}\|_{\infty} 
\le  
\sum_{\newatop{X\cap\Lambda\neq0:} 
              {\xi(X)=g}} 
 \|V_{X,\Lambda}\|_\infty 
\le  
\sum_{x\in g} 
\sum_{\newatop{X\cap\Lambda\neq0:} 
              {X\ni x}} 
 \|V_{X,\Lambda}\|_\infty 
\le  
|g| A 
\end{equation} 
The bound \eqref{sauto0} follows from $|g|\le(\Gamma_i+1)^d$. 
 
To prove \eqref{sint0} we first note that for $G\su\subset\cG_{\ge i}$, 
such that $G\cap\cG_{i}\neq\es$ and $(|G|,s)\neq(1,0)$, and 
$X\in\Upsilon_{\Lambda}(G,s)$ we have, recalling 
\eqref{esg0} and item~\ref{gent:dist} in definition~\ref{t:gentle},  
that $\tree(X)\ge\g_i$.  
Therefore by using (\ref{alb_iY}) we have  
\begin{equation} 
\label{prpr} 
\begin{array}{rl} 
{\displaystyle 
\inf_{X\in\Upsilon_{\Lambda}(G,s)}\tree(X)} & 
{\displaystyle 
 \ge 
 \frac{1}{4}\gamma_i 
 + 
 \frac{1}{4}\hat\cT_0(G,s) 
 + 
 \frac{1}{2} 
 \inf_{X\in\Upsilon_{\Lambda}(G,s)}\tree(X)}\\ 
& 
{\displaystyle 
 \ge 
 \frac{1}{4}\gamma_i 
 + 
 \frac{1}{4}\hat\cT_0(G,s) 
 + 
 \frac{1}{4} 
 \gamma_i\big[(|G|-1)\vee1\big]}\\ 
\end{array} 
\end{equation} 
where in the last step we used Lemma~\ref{t:infpi} in the case $|G|\ge 2$. 
Now, for $G$ and $s$ as above, remarking that 
$|G|\ge2$ implies $|G|-1\ge|G|/2$, we have, recalling  
$\gamma_i\ge\gamma_1$ and  
$\varepsilon=\exp\{-\alpha\gamma_1/8\}$ as 
in item~\ref{i:main3} of Theorem~\ref{t:mainceds}, 
\begin{equation} 
\label{pfsint0} 
\begin{array}{rcl} 
{\displaystyle 
\|\F^{(i,0)}_{G,s,\Lambda}\|_{\infty}} 
&\le& 
{\displaystyle 
 \sum_{X\in\Upsilon_{\Lambda}(G,s)}\|V_{X,\Lambda}\|_\infty 
 =  
 \sum_{X\in\Upsilon_{\Lambda}(G,s)}e^{\a\tree(X)}e^{-\a \tree(X)} 
                           \|V_{X,\Lambda}\|_\infty}\\ 
& \le & 
{\displaystyle 
 e^{ 
 -\frac{1}{4}\alpha\gamma_i 
 -\frac{1}{4}\alpha\hat\cT_0(G,s) 
 -\frac{1}{8}\alpha 
 \gamma_1|G|} 
 \sum_{X\in\Upsilon_{\Lambda}(G,s)}e^{\a\tree(X)} 
                           \|V_{X,\Lambda}\|_\infty}\\ 
& \le & 
{\displaystyle 
 \varepsilon^{|G|} 
 e^{-\frac{1}{4}\alpha\hat\cT_0(G,s)} 
 e^{-\frac{1}{4}\alpha\gamma_i} 
 \sup_{g\in\cG_{i}} 
 \sum_{\stackrel{X\ssu\bL}{\scriptscriptstyle \xi(X)\ni g}} 
     e^{\a\tree(X)}\|V_{X,\Lambda}\|_\infty}\\ 
& \le & 
{\displaystyle 
 \varepsilon^{|G|} 
 e^{-\frac{1}{4}\alpha\hat\cT_0(G,s)} 
 e^{-\frac{1}{4}\alpha\gamma_i} 
 \sup_{g\in\cG_{i}} 
 \sum_{x\in g} 
 \sum_{\stackrel{X\ssu\bL}{\scriptscriptstyle X\ni x}} 
     e^{\a\tree(X)}\|V_{X,\Lambda}\|_\infty}\\ 
& \le & 
{\displaystyle 
 \varepsilon^{|G|} 
 e^{-\frac{1}{4}\alpha\hat\cT_0(G,s)} 
 e^{-\frac{1}{4}\alpha\gamma_i} 
 (\Gamma_i+1)^dA}\\ 
\end{array} 
\end{equation} 
where we used the same bound as in \eqref{auto01}. 
The bound (\ref{sint0}) finally follows from item \ref{cc2} in 
the hypotheses of Theorem~\ref{t:rispot}. 
\qed 
 
\bpro{t:re} 
Let the hypotheses of Theorem~\ref{t:mainceds} 
be satisfied. 
Let also $\F^{(j,h)}_{G,s,\Lambda}$ satisfy the bound 
\eqref{recpot} for any $G\ssu\cG_{\ge j}$ with 
$G\cap\cG_{j}\neq\es$, any $s\ge 0$, and any \mbox{$h=0,\dots,j-1$}. 
Then, for each $\L\ssu\bL$, the cluster expansion in \eqref{polce} is 
absolutely convergent. 
Moreover, $\Ps^{(i,j)}_{g,\Lambda}$ and 
$\F^{(i,j)}_{G,s,\Lambda}$, 
as defined in (\ref{selfintij}) and (\ref{effiintij}), satisfy the bounds 
\eqref{recauto} and \eqref{recpot} 
for any $i>j\ge1$. 
\epro 
 
The proof of the inductive step in Proposition~\ref{t:re} is split in 
a series of Lemmata in which we understate the hypotheses of 
Proposition~\ref{t:re} itself to be satisfied. 
 
\blem{t:sacRj} 
For $R\in\cR_j$, let $\z_{R,\Lambda}$ be as defined in \eqref{acRj}. 
Then we have 
\be{sacRj} 
\|\z_{R,\Lambda}\|_{\infty} 
\le \prod_{(G,s)\in R} 
     \e^{|G|}\, e^{-(m_{j-1}-\delta_j)\hat{\cT}_{j-1}(G,s)} 
\end{equation} 
for any $\Lambda\subset\subset\bL$. 
\elem 
 
\Pro 
Recalling \eqref{intj}, the inductive hypotheses \eqref{recpot} 
implies that for each $G\ssu\cG_{\ge j}$, $G\cap\cG_j\neq\es$ and $s\ge 
0$ with $(|G|,s)\neq(1,0)$ 
\be{sintj} 
\|\F^{(j)}_{G,s,\Lambda}\|_{\infty} 
\le 
\sum_{h=0}^{j-1}  \e^{|G|} e^{-m_h \hat{\cT}_h (G,s)} 
\le  
j\, \e^{|G|}  e^{-m_{j-1} \hat{\cT}_{j-1} (G,s)} 
\end{equation} 
where we used that $m_h$, $\hat{\cT}_h$ 
are decreasing in $h$. 
Note that for $g\in\cG_{\ge j}$ and $s\ge 1$ we have, by recalling 
the inequality (\ref{alb_iY2}) and definition \eqref{esg0}, 
that 
$$ 
\hat{\cT}_h(g,s)\ge\frac{1}{4} 
                      \dis(\env(g),\dEs_s(g))> 
                         \frac{1}{4}\gamma_j 
\quad\quad h=0,\dots,j-1 
$$ 
On the other hand, for $G\ssu\cG_{\ge j}$, 
$|G|\ge 2$, there are $g,g'\in G$ with $\dis(g,g') > \g_j$. Hence, 
recalling (\ref{alb_iY2}) 
$$ 
\hat{\cT}_h(G,s) 
\ge\frac{1}{2}\cT(G) 
\ge\frac{1}{2}\g_j 
\ge\frac{1}{4}\g_j 
\quad\quad h=0,\dots,j-1 
$$ 
We thus conclude that for each $G\su\subset\cG_{\ge j}$ such that 
$(|G|,s)\neq(1,0)$, $j\ge 1$, we have 
\be{Tj>} 
\hat{\cT}_h(G,s) 
\ge\frac{1}{4}\g_j 
\quad\quad h=0,\dots,j-1 
\end{equation} 
 
Since $\F^{(j)}_{G,s,\Lambda}=0 $ if $(|G|,s)=(1,0)$, 
$m_{j-1}\g_j\ge 32j$ 
(see item \ref{conv:5} in Lemma~\ref{t:condsupp}), 
and $\e\in(0,1)$, from \eqref{sintj} 
we get the bound $\|\F^{(j)}_{G,s,\Lambda}\|_\infty\le 1$. 
Recalling definition \eqref{acRj} of the activity of a 
$j$--polymer $R$ and using the bound $|e^x-1|\le e^{|x|} |x|$ and 
\eqref{sintj}, we get 
\begin{eqnarray*} 
\|\z_{R,\Lambda}\|_{\infty} 
&\le& \prod_{(G,s) \in R}  ej  \, \e^{|G|} e^{-m_{j-1} \hat{\cT}_{j-1} (G,s)} 
\\ 
&\le& \prod_{(G,s) \in R} 
\e^{|G|}\, e^{-(m_{j-1}-\delta_j)\hat{\cT}_{j-1}(G,s)} 
\sup_{j \ge 0}\Big[ej\exp\big\{-\delta_j\frac{1}{4}\g_j\big\}\Big] 
\end{eqnarray*} 
where we used again \eqref{Tj>}. The bound \eqref{sacRj} follows 
since $\sup_{r\ge 0} \{e\,r\,e^{-r}\}=1$ 
and 
$\delta_j\g_j\ge 8j\ge 4j$, 
see item \ref{cc1} Lemma~\ref{t:condsupp}. 
\qed 
 
\blem{t:pettini} 
For $R\in\cR_j$, let 
\be{tz} 
{\tilde \z}_R := \e^{| R\rest_j |} 
\prod_{(G,s) \in R} 
\exp\big\{-\delta_j\hat{\cT}_{j-1}(G,s)\big\} 
\end{equation} 
Then 
\be{pettini} 
\sup_{g\in\cG_j} 
\sum_{\stackrel{R\in\cR_j} 
               {\scriptscriptstyle R\rest_j\ni g}} 
{\tilde \z}_R\,\exp\big\{|R\rest_j|\big\} 
\le 1 
\end{equation} 
\elem 
 
\Pro 
The above Lemma follows from the estimate in 
\cite[Appendix B]{[CasO]}, indeed the only needed ingredient is provided by 
Lemma~\ref{t:stepj}. Firstly we notice that from definition (\ref{tz}) 
we have 
$$ 
{\tilde \z}_R\, e^{|R\rest_j|}= 
 (e\varepsilon)^{|R\rest_j|} \prod_{(G,s)\in R} 
 \exp\big\{-\delta_j\hat{\cT}_{j-1}(G,s)\big\} 
$$ 
{}From item \ref{conv:4} in Lemma~\ref{t:condsupp} and 
Lemma~\ref{t:stepj} we get 
\begin{equation} 
\label{mer0.1} 
\sup_{g\in\cG_j} \sum_{\stackrel{G\ssu\cG_{\ge j}} 
                                {\scriptscriptstyle G\ni g}} 
\sum_{s=0}^\infty \id_{(|G|,s)\neq(1,0)} 
\exp\big\{-\delta_j\hat{\cT}_{j-1}(G,s)\big\} 
\le K_j(\delta_j)=:\tilde{K}_j 
\end{equation} 
On the other hand from items \ref{conv:0} and \ref{conv:3} in 
Lemma~\ref{t:condsupp} we easily get 
\be{ipettini} 
e^{{\tilde K}_j}  \le \frac {1}{ e\,\e (2 -e\,\e)} 
\;\;\;\;\;\;\textrm{for all}\;\;j\ge 1 
\end{equation} 
 
Now, by using (\ref{ipettini}) and item \ref{conv:0} in Lemma~\ref{t:condsupp} 
we can indeed perform the estimate in \cite[Appendix B]{[CasO]} 
to obtain 
\be{bpettini} 
\sup_{g\in\cG_j} 
\sum_{\stackrel{R\in\cR_j} 
               {\scriptscriptstyle R\rest_j\ni g}} 
     {\tilde \z}_R\: e^{|R\rest_j|} 
\le 
e\varepsilon{\tilde K}_j\Big[1+ 
                        \frac{e^{{\tilde K}_j}-1} 
                             {1+(e\varepsilon)^2e^{{\tilde K}_j} 
                             -2e\varepsilon e^{{\tilde K}_j}}\Big] 
\le 1 
\end{equation} 
where the last inequality follows from items \ref{conv:0} 
and \ref{conv:3} of Lemma~\ref{t:condsupp} by elementary computations. 
\qed 
 
The bound \eqref{pettini} allows us to justify the cluster expansion 
in \eqref{polce}. We are now indeed ready to apply the abstract 
theory developed in \cite{[KP]}. 
 
\blem{t:KP} 
For $R\in\cR_j$, let ${\tilde\z}_R$ as in \eqref{tz} 
and, for $\ul{R}\in\ul{\cR}_j$, 
set ${\tilde\z}_{\ul{R}}:=\prod_{R\in\ul{R}} {\tilde\z}_R$. 
Then, recalling the incompatibility $\inc_j$ has been defined below 
(\ref{cRiinc}), for each $S\in \cR_j$ we have 
\be{e:KP} 
\sum_{\stackrel{\ul{R}\in\ul{\cR}_j} 
               {\scriptscriptstyle \ul{R}\,\inc_j\,S}} 
\lmo \f_T(\ul{R})\rmo\,  {\tilde\z}_{\ul{R}}  \le | S \rest_j | 
\end{equation} 
\elem 
 
\Rem Since, by (\ref{aa3}) 
$m_{j-1}-\delta_j\ge m_j-\delta_j\ge3\delta_j\ge\delta_j$, 
from Lemmata 
\ref{t:distre}, \ref{t:sacRj}, \ref{t:KP}, and (\ref{tz}) it 
follows for each $\L\ssu\bL$ the cluster expansion in \eqref{polce} is 
absolutely convergent if the hypotheses of Theorem~\ref{t:mainceds} hold. 
This proves the first claim in Proposition~\ref{t:re}. 
 
\medskip 
\noi{\it Proof of Lemma~\ref{t:KP}.\/}\ \ 
For each $S\in\cR_j$ we have the bound 
$$ 
\sum_{\stackrel{R \in \cR_j} 
               {\scriptscriptstyle R \,\inc_j\,S}} 
{\tilde \z}_R \, e^{| R\rest_j|} \le 
\sum_{g\in S\rest_j } 
\sum_{\stackrel{R\in\cR_j} 
               {\scriptscriptstyle R\rest_j\ni g}} 
{\tilde \z}_R \, e^{| R\rest_j| } 
\le |S \rest_j | 
$$ 
where we applied Lemma~\ref{t:pettini}. 
The bound \eqref{e:KP} now follows from the Theorem in \cite{[KP]}  
by choosing there $a(R)=|R\rest_j|$. 
\qed 
 
We can now estimate the self interaction due the integration on scale 
$j$. 
 
\blem{t:sselfij} 
Let $g\in\cG_i$ and $\Ps^{(i,j)}_{g,\Lambda}$ as defined in 
\eqref{selfintij}. Then for each $i\ge j+1$ 
\be{sselfij} 
\|\Psi^{(i,j)}_{g,\Lambda}\|_{\infty} 
\le A_i 
\end{equation} 
for any $\Lambda\subset\subset\bL$. 
\elem 
 
\Pro 
Recalling \eqref{tz}, by 
using Lemmata~\ref{t:distre} and \ref{t:sacRj}, we get 
\begin{equation} 
\label{sij1} 
\begin{array}{rcl} 
{\displaystyle 
\|\Ps^{(i,j)}_{g,\Lambda}\|_{\infty}} 
& \le & 
{\displaystyle 
 \e \sum_{\ul{R}\in \ul{\cR}_j(g,0)} \lmo\f_T(\ul{R})\rmo 
 \,{\tilde\z}_{\ul{R}}\, 
 \prod_{R\in\ul{R}}\;\prod_{(H,u)\in R} 
 e^{-(m_{j-1}-2\delta_j){\hat\cT}_{j-1}(H,u)}}\\ 
& \le & 
{\displaystyle 
 \vphantom{\Bigg\{} 
 \e \sum_{h\in \cG_j} 
 \sum_{\stackrel{\ul{R}\in \ul{\cR}_j (g,0)} 
                {\scriptscriptstyle \ul{R}\rest_j\ni h}} 
 \lmo \f_T(\ul{R})\rmo 
 \,{\tilde\z}_{\ul{R}}\, 
 \prod_{R\in\ul{R}}\;\prod_{(H,u)\in R} 
 e^{-(m_{j-1}-2\delta_j){\hat\cT}_{j-1}(H,u)}}\\ 
& \le & 
{\displaystyle 
 \e \sum_{h\in \cG_j} 
 e^{-(m_{j-1}-2\delta_j)\disuno(g,h)/2} 
 \sup_{h\in \cG_j} 
 \sum_{\stackrel{\ul{R}\in \ul{\cR}_j} 
                {\scriptscriptstyle \ul{R}\rest_j\ni h}} 
 \lmo \f_T(\ul{R})\rmo 
 \,{\tilde\z}_{\ul{R}}}\\ 
\end{array} 
\end{equation} 
where we used (\ref{dgg'}). 
 
We next observe that for $h\in\cG_j$,  by the notion of 
$j$--incompatible $j$--polymers, we have that $\ul{R} \rest_j\ni h$ implies 
$\ul{R}\,\inc_j\,(h,0)$. Therefore, by Lemma~\ref{t:KP}, 
\be{sij2} 
\sup_{h\in \cG_j} 
\sum_{\stackrel{\ul{R}\in \ul{\cR}_j} 
               {\scriptscriptstyle \ul{R}\rest_j\ni h}} 
\lmo \f_T(\ul{R})\rmo {\tilde\z}_{\ul{R}} 
\le 1 
\end{equation} 
Finally, 
\begin{equation} 
\label{sij3} 
\begin{array}{rl} 
{\displaystyle 
 \sum_{h\in \cG_j} 
 e^{-\frac{m_{j-1}-2\delta_j}{2}\disuno(g,h)} \le }& 
{\displaystyle 
 \sum_{y\in\bL} 
 e^{-\frac{m_{j-1}-2\delta_j}{2}\disuno(y,B_{\G_i})}}\\ 
\le & 
{\displaystyle 
 \[ 2 \( 2\G_i+\g_j\) +1 \]^d 
 +\!\!\! 
  \sum_{\newatop{y\in\bL:}{y\not\in B_{2\Gamma_i+\gamma_j}}} 
 \!\!\!\! 
 e^{-\frac{m_{j-1}-2\delta_j}{2}\dis(y,B_{\G_i})}}\\ 
\le & 
{\displaystyle 
 4^d [\G_i +\g_j]^d + K\((m_{j-1}-2\delta_j)/2\)\, 
 e^{-\frac{m_{j-1}-2\delta_j}{4}\g_j}}\\ 
\end{array} 
\end{equation} 
where we used Lemma~\ref{t:elbound}. 
 
Noticing that item~\ref{seq:con-dist} in Definition~\ref{t:seq} 
implies $\g_j \le \G_{j+1}\le \G_i$ and recalling item \ref{conv:6} in 
Lemma~\ref{t:condsupp}, the bound \eqref{sselfij} follows. 
\qed 
 
The recursive estimate on the effective interaction due the 
integration on scale $j$ requires now only a little extra effort. 
Indeed, the proof of Proposition~\ref{t:re} is concluded by the 
following Lemma. 
 
\blem{t:spotij} 
Let $G\ssu\cG_{\ge i}$, $G\cap \cG_i\neq\es$, $s\ge 0$ 
and $\F^{(i,j)}_{G,s,\Lambda}$ as defined in 
\eqref{effiintij}. 
Then for each $i\ge j+1$ 
\be{spotfij} 
\|\F^{(i,j)}_{G,s,\Lambda}\|_{\infty} 
\le 
\e^{|G|}\, e^{-(m_{j-1}-4\delta_j)\hat{\cT}_j(G,s)} 
\end{equation} 
for any $\Lambda\subset\subset\bL$. 
\elem 
 
\Pro 
Let $g \in G\cap \cG_i$;  
recall definition (\ref{alb_iY}), by applying 
(\ref{dgg'}), Lemmata~\ref{t:distre}, \ref{t:sacRj}, and  
using the same bounds as in \eqref{sij1} we get 
\begin{equation} 
\label{pij1} 
\begin{array}{rcl} 
{\displaystyle 
\|\F^{(i,j)}_{G,s,\Lambda}\|_{\infty}} 
& \le & 
{\displaystyle 
 \e^{|G|} e^{-(m_{j-1}-3\delta_j)\hat{\cT}_j(G,s)} 
 \sum_{h\in \cG_j} 
 e^{-\delta_j\dis(g,h)/2} 
 \sup_{h\in \cG_j} 
 \sum_{\ul{R}\in \ul{\cR}_j:\;\ul{R}\rest_j\ni h} 
 \lmo \f_T(\ul{R})\rmo 
 {\tilde\z}_{\ul{R}}}\\ 
&\le & 
{\displaystyle 
 \e^{|G|} 
 e^{-(m_{j-1}-4\delta_j)\hat{\cT}_j(G,s)-\delta_j\gamma_i/4} 
 \[4^d[\G_i+\g_j]^d+K(\delta_j/2)\, e^{-\delta_j\g_j/4}\]}\\ 
\end{array} 
\end{equation} 
where we used \eqref{Tj>} and \eqref{sij2}, and argued as in  
\eqref{sij3}. 
Recalling the bounds \ref{conv:7} and \ref{conv:8} in 
Lemma~\ref{t:condsupp} the estimate \eqref{spotfij} is proven. 
\qed 
 
With the proof of this Lemma it is also completed the proof of 
Proposition~\ref{t:re}. 
We finally show how to get Theorem~\ref{t:mainceds} from 
\eqref{recauto} and \eqref{recpot}. 
 
\noi{\it Proof of Theorem~\ref{t:mainceds}.\/}\ \ 
Item~\ref{i:main1}: 
equation (\ref{ZL}) has been formally obtained in Section \ref{s:algebra}; 
the absolute convergence, uniform with respect to $\Lambda$, of the 
series involved in (\ref{ZL}) follows from Proposition~\ref{t:re}. 
Item~\ref{i:main2} follows immediately from the remarks below 
definitions (\ref{ZG}) and (\ref{accrj}). 
Item~\ref{i:main3}: 
to prove the bound (\ref{eautopot}) 
we recall (\ref{Ug}), (\ref{ZG}), Theorem~\ref{t:bounds} and 
$S:=\sup_{x\in\bL}|\cS_x|$ to get 
\begin{equation} 
\label{prro} 
\|\log Z_{g,\Lambda}^{(j)}\|_{\infty} 
\le 
|g|(\log S+\|U\|)+\sum_{h=0}^{j-1}A_h 
\end{equation} 
which implies the thesis. 
Finally, to get the bound (\ref{polbound}) 
we have to use equation (\ref{sacRj}) in definition (\ref{accrj}), 
the obvious fact that $m_{j-1}-\delta_j>m_j=m_{j-1}-4\delta_j$ 
(see (\ref{cc71})), 
(\ref{alb_iY2}), and the fact that $m_j\ge\alpha/8$, which 
follows from (\ref{cc71}) and item~\ref{cc17} in Lemma~\ref{t:condsupp}. 
\qed 
 
\sezione{Proof of the main theorems}{s:prot} 
\par\noindent 
First of all we show that Theorem~\ref{t:rispot} is 
a consequence of cluster expansion stated in 
Theorem~\ref{t:mainceds}. 
 
\noi{\it Proof of Theorem~\ref{t:rispot}.\/}\ 
Recalling (\ref{costa}) and 
the notation introduced in Section~\ref{s:algebra}, 
for $\Lambda,X\subset\subset\bL$ 
we set 
\begin{equation} 
\label{pt0} 
\Psi_{X,\Lambda,0}:= 
\left\{ 
\begin{array}{ll} 
{\displaystyle 
 V_{X,\Lambda}} 
& \;\;\;\;\textrm{if}\;\diaminfinito(X)\le\varrho,\; \xi(X)=\emptyset, 
\textrm{ and } X\cap\Lambda\neq\emptyset\\ 
0 &\;\;\;\;\textrm{otherwise}\\ 
\end{array} 
\right. 
\end{equation} 
and 
\begin{equation} 
\label{pt1} 
\Phi_{X,\Lambda,0}:= 
\left\{ 
\begin{array}{ll} 
{\displaystyle 
 V_{X,\Lambda}} 
& \;\;\;\;\textrm{if}\;\diaminfinito(X)>\varrho,\,\xi(X)=\emptyset, 
\textrm{ and } X\cap\Lambda\neq\emptyset\\ 
0 &\;\;\;\;\textrm{otherwise}\\ 
\end{array} 
\right. 
\end{equation} 
Note that the families $\{\Psi_{X,\Lambda,0},\,\Lambda\subset\subset\bL\}$  
and $\{\Phi_{X,\Lambda,0},\,\Lambda\subset\subset\bL\}$  
are $(X,\emptyset)$--compatible. Moreover, for $j\ge 1$ 
\begin{equation} 
\label{ppsip} 
\begin{array}{l} 
\Psi_{X,\Lambda,j}:={\displaystyle 
                     \sum_{g\in\cG_j:\,\Es_0(g)=X} 
                     \log Z_{g,\Lambda}^{(j)}\vphantom{\Bigg(}}\\ 
\Phi_{X,\Lambda,j}:={\displaystyle 
                     \sum_{\newatop{\ul{R}\in\ul{\cR}_j:} 
                                 {\ul{R}\rest_{\ge j+1}=\emptyset,\, 
                                  \supp\ul{R}=X}} 
                   \varphi_T(\ul{R})\zeta_{\ul{R},\Lambda}\vphantom{\Bigg(}}\\ 
\end{array} 
\end{equation} 
We finally set 
$\Psi_{X,\Lambda}:=\sum_{j=0}^{\varkappa}\Psi_{X,\Lambda,j}$ and 
$\Phi_{X,\Lambda}:=\sum_{j=0}^{\varkappa}\Phi_{X,\Lambda,j}$, 
recall $\varkappa$ has been introduced in item~\ref{i:main1} of  
Theorem~\ref{t:mainceds}.  
 
{}From equation (\ref{ZL}) and the previous definitions we have that 
the identity (\ref{tm1}) holds. 
On the other hand, from Condition~\ref{t:ipotesi}, 
the $(\Es_0(g),\emptyset)$--compatibility of $Z_{g,\Lambda}$, 
and the $(\supp\ul{R},\proj{\ul{R}\rest_{\ge j+1}})$--compatibility of 
$\zeta_{\ul{R},\Lambda}$ we easily get that item~\ref{p:tm2} holds true. 
 
Now, from (\ref{pt0}) and (\ref{ppsip}) it follows that if  
$\diaminfinito(X)>\varrho$ and  
$\nexists g\in\cG_{\ge 1}$ such that $\Es_0(g)=X$ then 
$\Psi_{X,\Lambda}=0$. Moreover, recalling item~\ref{gent:cas} 
in Definition~\ref{t:gentle}, for each $x\in\bL$ we get 
\begin{equation} 
\label{pirro} 
\sum_{X\ni x} 
\sup_{\Lambda\subset\subset\bL} 
\|\Psi_{X,\Lambda}\|_{\infty} 
\le 
\sum_{X\ni x} 
\sup_{\Lambda\subset\subset\bL} 
\Big[ 
 \|\Psi_{X,\Lambda,0}\|_\infty 
 + 
  \sum_{j=1}^{k_x}\sum_{\newatop{g\in\cG_j:}{\Es_0(g)\ni x}} 
    \|\log Z_{g,\Lambda}^{(j)}\|_{\infty} 
\Big] 
\end{equation} 
By exploiting (\ref{ip2}) in Condition~\ref{t:ipotesi}, 
the first term on the right--hand side of (\ref{pirro}) 
can be easily bounded as follows 
\begin{equation} 
\label{pirro2} 
\sum_{X\ni x} 
\sup_{\Lambda\subset\subset\bL} 
\|\Psi_{X,\Lambda,0}\|_{\infty} 
\le 
 \sum_{X\ni x} 
 \sup_{\Lambda\subset\subset\bL} 
 \|V_{X,\Lambda}\|_{\infty} 
\le 
A 
\end{equation} 
To bound the second term on the right--hand side of (\ref{pirro}) we note that 
$\big|\{g\in\cG_j:\,\Es_0(g)\ni x\}\big|\le[\Gamma_j+1+2\vartheta_j]^d$. 
Hence the bound (\ref{tm3}), which completes the proof of item~\ref{p:tm3}, 
follows from the above inequality, (\ref{eautopot}), 
(\ref{pirro}), and (\ref{pirro2}). 
 
In order to prove item~\ref{p:tm4} 
let us first show that for $G\subset\subset\cG_j$ and $s\ge0$, if 
$(|G|,s)\neq(1,0)$ we have 
\be{a1.1} 
\hat\cT_{j-1}(G,s)\ge\frac{1}{12}\diaminfinito\big(\Es_s(G)\big) 
\end{equation} 
It is interesting to remark that the bound (\ref{a1.1}) might fail 
if it were $G\subset\subset\cG_{\ge j}$ and $G\cap\cG_{\ge j+1}\neq\emptyset$. 
If $|G|=1$ then $G=\{g\}$ for some $g\in\cG_j$; by 
recalling \eqref{esg0}, \eqref{alb_iY2} 
we get, since $s\ge 1$ and $\vartheta_j>3\Gamma_j$, 
$$ 
\hat\cT_{j-1}(\{g\},s)=\frac{1}{4}\dis\(\env(g),\dEs_s(g)\) 
\ge 
\frac{1}{4}(\vartheta_j+s) 
\ge 
\frac{1}{12}(2\vartheta_j+2s+\Gamma_j) 
\ge 
\frac{1}{12}\diaminfinito\(\Es_s(g)\) 
$$ 
Let, now, $|G|\ge 2$ and $s=0$. 
Recall $\vartheta_j<2\gamma_j$ and $\cT(G)\ge\gamma_j>2\Gamma_j$. 
By applying (\ref{alb_iY2}) we get  
\begin{equation*} 
\hat\cT_{j-1}(G,s) 
  \ge \frac{1}{2}\cT(G)\ge 
        \frac{1}{3}\gamma_j+\frac{1}{6}\cT(G)\ge 
        \frac{1}{6}\vartheta_j+\frac{1}{12}\diaminfinito\big(\env(\proj{G})\big) 
        \ge 
        \frac{1}{12}\diaminfinito\big(\Es_0(\proj{G})\big) 
\end{equation*} 
Finally, in the case $|G|\ge 2$ and $s\ge 1$ by (\ref{alb_iY2}) 
\begin{equation*} 
\hat\cT_{j-1}(G,s) 
  \ge \frac{1}{6}\cT(G)+\frac{1}{6}(\vartheta_j+s)\ge 
      \frac{1}{12}\diaminfinito\big(\env(\proj{G})\big) 
        +\frac{1}{12}(2\vartheta_j+2s) 
  \ge\frac{1}{12}\diaminfinito\big(\Es_s(\proj{G})\big) 
\end{equation*} 
 
{}From (\ref{a1.1}) we get that, given $X\subset\subset\bL$, 
for any $\ul{R}\in\ul{\cR}_j$ 
such that $\supp\ul{R}=X$ and $\ul{R}\rest_{\ge j+1}=\emptyset$ we have 
\begin{equation} 
\label{ppio1} 
\sum_{R\in\ul{R}}\sum_{\newatop{(G,s)\in R:} 
                               {(|G|,s)\neq(1,0)}} 
\hat\cT_{j-1}(G,s)\ge\frac{1}{12}\diaminfinito(X) 
\end{equation} 
Furthermore, given $g\in\cG_j$ and $x\in\bL$, 
for any $\ul{R}\in\ul{\cR}_j$ 
such that $\supp\ul{R}\ni x$, $\ul{R}\rest_{j}\ni g$, and 
$\ul{R}\rest_{\ge j+1}=\emptyset$ we have that 
the left--hand side of (\ref{ppio1}) is bounded from below by 
$\disinfinito(x,g)/12$. 
Recalling (\ref{tz}), by applying 
Lemma~\ref{t:sacRj}, and noticing that 
$m_{j-1}-2\delta_j\ge m_j$, we have that for each $x\in\bL$ 
\begin{equation} 
\label{plot} 
\begin{array}{l} 
{\displaystyle 
\sum_{X\ni x} 
e^{q\alpha\diaminfinito(X)} 
\sup_{\Lambda\subset\subset\bL} 
\|\Phi_{X,\Lambda}\|_{\infty} 
\le 
\sum_{X\ni x} 
e^{q\alpha\diaminfinito(X)} 
\,\bigg[ 
\sup_{\Lambda\subset\subset\bL} 
\|\Phi_{X,\Lambda,0}\|_{\infty} 
}\\ 
\\ 
{\displaystyle 
\phantom{merd} 
+ 
\sum_{j\ge 1}\sum_{\newatop{\ul{R}\in\ul{\cR}_j:} 
                           {\supp\ul{R}=X,\,\ul{R}\rest_{\ge j+1}=\emptyset}} 
\!\!\!\!\!\!\!\big|\varphi_T(\ul{R})\big| 
\, 
\exp\bigg\{-m_j\sum_{R\in\ul{R}}\sum_{\newatop{(G,s)\in R:} 
                                      {(|G|,s)\neq(1,0)}} 
    \hat\cT_{j-1}(G,s)\bigg\} 
\cdot\tilde\zeta_{\ul{R}}\bigg]}\\ 
\end{array} 
\end{equation} 
Recalling (\ref{pt1}), 
the first term on the right--hand side of (\ref{plot}) 
can be bounded as follows 
$$ 
\begin{array}{rcl} 
{\displaystyle 
\!\! 
\sum_{X\ni x} 
e^{q\alpha\diaminfinito(X)} 
\!\! 
\sup_{\Lambda\subset\subset\bL} 
\|\Phi_{X,\Lambda,0}\|_{\infty}} 
&\!\!\!\le& 
{\displaystyle 
 \sum_{\newatop{X\ni x:} 
               {\diaminfinito(X)>\varrho}}e^{q\alpha\tree(X)} 
\sup_{\Lambda\subset\subset\bL} 
 \|V_{X,\Lambda}\|_{\infty}}\\ 
&\!\!\!\le& 
{\displaystyle 
 e^{-(1-q)\varrho\alpha} 
 \sum_{X\ni x} 
 e^{\alpha\tree(X)} 
\sup_{\Lambda\subset\subset\bL} 
 \|V_{X,\Lambda}\|_{\infty}} 
\le 
{\displaystyle 
 A 
 e^{-(1-q)\varrho\alpha} 
 \le e^{-\alpha}}\\ 
\end{array} 
$$ 
where we used $\tree(X)\ge\diaminfinito(X)$, 
Condition~\ref{t:ipotesi}, and definitions (\ref{costa}). 
 
Recall $q=2^{-5}3^{-2}$, 
by using (\ref{ppio1}), the remark below it, (\ref{Tj>}), 
and $m_j\ge\alpha/8$ we get, by simple computations, 
that the second term on the right--hand 
side of (\ref{plot}) can be bounded by 
\begin{equation} 
\label{plota} 
e^{-m_j\gamma_1/36}\, 
\sum_{X\ni x}\sum_{j\ge 1}  
 e^{-m_j\gamma_j/18} 
\sum_{g\in\cG_j}  
 e^{-(m_j/36)\disinfinito(x,g)} 
\sum_{\newatop{\ul{R}\in\ul{\cR}_j:\,\ul{R}\rest_j\ni g} 
              {\supp\ul{R}=X,\,\ul{R}\rest_{\ge j+1}=\emptyset}} 
|\varphi_T(\ul{R})|\tilde\zeta_{\ul{R}} 
\end{equation} 
which, in turn, by Lemma~\ref{t:KP} is bounded by 
\begin{equation} 
\label{pelota} 
\begin{array}{rcl} 
{\displaystyle 
e^{-q\alpha\gamma_1}\, 
\sum_{j\ge 1} e^{-m_j\gamma_j/18} 
\sum_{g\in\cG_j} e^{-(m_j/36)\disinfinito(x,g)}} 
&\le& 
{\displaystyle 
e^{-q\alpha\gamma_1}\, 
\sum_{j\ge 1} e^{-32j/18} 
\sum_{y\in\bL} e^{-\alpha q\disinfinito(y,x)}}\\ 
&\le& 
{\displaystyle 
e^{-q\alpha\gamma_1}\, 
\frac{e^{-16/9}}{1-e^{-16/9}} 
\,K\Big(\frac{q\alpha}{d}\Big) 
\le 
e^{-q\alpha\gamma_1}\, 
\,K\Big(\frac{q\alpha}{d}\Big)}\\ 
\end{array} 
\end{equation} 
where we used item~\ref{conv:5} in Lemma~\ref{t:condsupp}, 
Lemma~\ref{t:elbound}, and the bound  
$\disinfinito(y,x)\ge\disuno(y,x)/d$. 
Recalling the function $K$ has been defined in (\ref{dKm}),  
we have proven the bound (\ref{tm2}) 
which completes the proof of the Theorem. 
\qed 
 
Theorem~\ref{t:decay} follows from Theorem~\ref{t:rispot}  
by the combinatorial techniques in \cite{[BCOalb]}. We are, indeed,  
in a situation analogous to \cite[Rem.\ \incr]{[BCOalb]}  
and it is not difficult to check that items \ref{i:ris2} and  
\ref{i:ris1} in the hypotheses of Theorem~\ref{t:decay}  
on the geometry of the supports of the local functions  
$f_1,\dots,f_n$ imply that Lemma~\iincr ~in \cite{[BCOalb]}, 
which yields the bound (\ref{adcs}), holds.  
 

\end{document}